\documentclass[twocolumn,superscriptaddress,showpacs,10pt]{revtex4}
\usepackage[titletoc,title,toc]{appendix}
\usepackage{graphicx}
\usepackage{amssymb}
\usepackage{epstopdf}
\usepackage{amsmath}
\usepackage{hyperref}
\usepackage{mathrsfs}
\usepackage{varioref}
\usepackage[active]{srcltx}
\usepackage{tikz}
\usepackage{lmodern}
\usetikzlibrary{decorations.pathmorphing}
\usepackage{fancybox}
\tikzset{snake it/.style={decorate, decoration=snake}}
\labelformat{section}{Section #1} 
\labelformat{subsection}{Section #1} 
\labelformat{equation}{Eq.(#1)} 
\labelformat{figure}{Fig.~#1} 
\labelformat{subfigure}{Fig.~\thefigure#1} 
\labelformat{table}{Tab.~#1} 
\newcommand{\be}{\begin{equation}}
\newcommand{\ee}{\end{equation}}
\newcommand{\bea}{\begin{eqnarray}}
\newcommand{\eea}{\end{eqnarray}}

\begin{document}

\title{Information retrieval from black holes}

\author{Kinjalk Lochan}%
\email{kinjalk@iucaa.in}%
\author{Sumanta Chakraborty}%
\email{sumanta@iucaa.in}%
\author{T. Padmanabhan}%
\email{paddy@iucaa.in}
\affiliation{IUCAA, Post Bag 4, Ganeshkhind,\\
Pune University Campus, Pune 411 007, India}
\begin{abstract}
 It is generally believed that, when matter collapses to form a black hole, the complete information about the initial state of the matter cannot be retrieved by future asymptotic observers, through local measurements. This is contrary to the expectation from a unitary evolution in quantum theory and leads to (a version of) the black hole information paradox. Classically, nothing else, apart from mass, charge and angular momentum is expected to be revealed to such asymptotic observers after the formation of a black hole. Semi-classically, black holes evaporate after their formation through the Hawking radiation. The dominant part of the radiation is expected to be thermal and hence one cannot know anything about the initial data from the resultant radiation. However, there can be sources of distortions which make the radiation  non-thermal. Although the distortions are not strong enough to make the evolution unitary, these distortions  carry some part of information regarding the in-state. In this work, we 
show how one can decipher the information about the in-state of the field from these distortions.
 We show that the distortions of a particular kind --- which we call {\it non-vacuum distortions} --- can be used to \emph{fully} reconstruct the initial data. The asymptotic observer can do this operationally by measuring certain well-defined observables of the quantum field at late times. We demonstrate that a general class of in-states  encode all their information content in the correlation of late time out-going modes. Further, using a $1+1$ dimensional CGHS model to accommodate back-reaction self-consistently, we show that  observers  can also infer and track the information content about the initial data, during the  course of evaporation, unambiguously. Implications of such information extraction are discussed.
\end{abstract}
\maketitle
\section{Introduction : Black hole information paradox}

Evaporation of black holes, for many decades, has caused conceptual discomfort for the otherwise very successful quantum theory. The most basic and fundamental feature of the standard quantum theory, namely unitary evolution, is seemingly threatened if one tries extrapolating the results obtained at the semi-classical level \cite{Hawking:1974sw}. Insights of Bekenstein \cite{Bekenstein:1972tm,Unruh:1983ms}, suggested that the black holes must have an entropy  proportional to the area of their event horizons, for the second law of thermodynamics to work. 
Hawking \cite{Hawking:1974sw} proved that quantum effects could lead to the evaporation of the black hole which involves (i) a radiation of positive energy, (nearly) thermal spectrum of particles which an asymptotic observer can detect and (ii)a flux of negative energy flowing into the black hole decreasing its mass. So the mass lost by the black hole  appears in the form of energy of the thermal radiation. Although this effect completes the thermodynamic description of black holes, yet such a process, together with other properties of black holes, appears to violate the standard unitary quantum mechanics \cite{Mathur:2009hf}. 

The particles in the outgoing flux, received by the asymptotic observer, remain entangled with the particles in the in-going flux. The resulting Hawking radiation is thermal, precisely because we trace over the modes which entered the horizon. By such a process the black hole shrinks, losing the mass in the form of Hawking radiation. However, once the black hole completely evaporates by this process, there is an apparent paradox. At the end, there is nothing left for the outgoing particles to remain entangled with; yet they are in a mixed state since at no stage of the evaporation their entanglement with the interior modes was explicitly broken. This process, wherein a pure state evolves into a mixed state, is contrary to the standard unitary quantum evolution. 

There is also a related issue of the information content of the matter which had undergone the collapse to form the black hole or even matter which falls into the black hole after it is formed. No-hair conjectures \cite{MTW} suggest that no other information, apart from the mass, charge and angular momentum, of the matter that enters the event horizon can be available to the outside observers, at any stage. Therefore, all other information about matter crossing the horizon would end up on the singularity and get destroyed. Thus, all the information about the initial state of matter which is falling into the horizon (other than those captured by mass, charge or angular momentum) is not coded in the Hawking radiation and is not available to the asymptotic future observers. This part of information, which ended up in singularity, is lost forever. Such a situation seems to require  non-unitary evolution \cite{Mann:1990ci}. 

An initial resolution of the paradox stemmed from the idea that we might be making error in trusting the semi-classical Hawking process all the way to the complete evaporation of the black hole. In principle, when the black hole is large enough, semi-classical description should work fine. But as the black hole becomes smaller and smaller, the curvature at horizon begins to rise and at very high curvatures the quantum nature of gravity must become important, and the semi-classical approximation must break down. Therefore, quantum gravity --- rather than the semi-classical physics --- should govern the final  moments of the black hole evaporation. One then expects an overriding correction to the semi-classical description ${\cal O}(l_{\text P}^2)$ which makes it  non-thermal, and only becomes dominant when the black hole becomes of the Planck size. It must be noted that there  exist many other sources of distortions to the thermal Hawking radiation \cite{Visser:2014ypa}, apart form the 
quantum gravity induced corrections. These non-thermal corrections can, in principle, store some information. However, it can be shown \cite{Mathur:2009hf} that, since all such correction terms are sub-dominant in nature, none of these can help in making the theory unitary. Only corrections of ${\cal O}(1)$ can provide a possibility for unitary description, and we could identify no distortions  of that kind. Thus, in its new avatar the paradox seems more robust as far as restoring unitarity to the quantum evolution is concerned.

In the literature, there are many different proposals to handle this issue. There are suggestions advocating radical modification to the unitary quantum theory itself, to accommodate non-unitary processes \cite{Modak:2014vya}. Such modifications to the unitary quantum theory have also been argued for, using some other conceptual considerations \cite{Adler:1995iv,Bassi:2003gd}. However, these non-unitary quantum evolution models can also be applied to many other physical scenarios \cite{Bassi:2012bg} where the predictions will be at variance from the standard unitary theory, constraining the models. There are also suggestions that the black hole evaporation must halt at the Planck level and leave behind a Planck size remnant at the end of the process. However, irrespective of the size, mass and other classical features of the black hole, formed initially, the end product always has to be a Planck size remnant. This remnant should house all the information, which the out-going modes lack, in order to 
completely specify the state. Thus the complete description of a remnant and the Hawking radiation should be a pure state. This looks like a viable option. Still, it is not clear how   a Planck size remnant could accommodate the vast landscape of varying initial configurations which could have formed the initial black hole. Other interesting suggestions  include pinching of the spacetime \cite{Mathur:2008kg} which could, in principle, restore faith in the essential tenets of both classical gravity and the quantum theory. However, the implication of such  pinching effects for other types of horizons (e.g. Rindler, de Sitter) remains to be understood satisfactorily.

So we can summarize the crux of the information paradox as follows. When the black hole evaporates completely without leaving any remnant behind, one is justified in assuming that the entire information content of the collapsing body gets either destroyed or must be encoded in the resulting radiation. However, remnant radiation in this process is (dominantly) thermal, which is thermodynamically prohibited to contain much of the information and also incapable of making the theory unitary. Therefore, most of the information content of the matter which made the black hole in the first place is not available to the future asymptotic observers. 

In this work, we argue that this version of the paradox --- concerning the information content of the initial data ---  stems from a hybrid quantum/classical analysis of a process which is  fully quantum mechanical in nature. That is, it arises from an artificial division between a quantum test field and  the classical matter which collapses to form a black hole. When an event horizon is formed, the quantum field residing in its vacuum state at the beginning of collapse, gradually gets populated, erasing the black hole through a negative energy flux into the horizon with a corresponding positive energy flux appearing at infinity as thermal radiation \cite{Chakraborty:2015nwa,Singh:2014paa}. However, the matter which forms a black hole in the first place, is also fundamentally quantum mechanical in nature and should follow a quantum evolution. This, we believe, holds the key to the resolution of the paradox. We expect that the classical description to be true, at lowest order, leading to formation of an event 
horizon. However, the information that the collapsing material was inherently quantum mechanical in nature (e.g. a coherent state of the field which is collapsing) should not be completely ignored in studying this process. The matter which forms the black hole, if treated quantum mechanically, will populate its modes at future asymptotia non-thermally, in a manner which depends on its initial state. In this paper, we  demonstrate the presence of this effect at the semi-classical level. The result indicates that the no-hair theorems will be superseded  at the full quantum gravity level. 

Previously it has been shown that the particle content of the ingoing field modes makes the resulting Hawking radiation to be supplemented by a stimulated emission. Therefore, the radiation profile becomes non-thermal and thus capable of storing information. There have been studies 
(see e.g.,\cite{Wald:1976ka,Panangaden:1977pc,Sorkin:1986zj,Audretsch:1992py,Page:1993wv,Schiffer:1993ai,Muller:1993ng,Adami:2004uq}) regarding information content of corrected spectrum, form the point of view of information theory (Von Neumann entropy, channel capacity, etc.). We, however, do not commit to a particular specification of the information content but concentrate on the possibility of explicit reconstruction of the initial state of the field from the resultant radiation in a collapse process. Neither do we make an attempt to restore unitarity by such a stimulated emission process. Our focus will be to reconstruct the initial data to the extent possible when the Hawking radiation has a non-thermal part. We show that whenever a field which enters the horizon is in a non-vacuum configuration, it ends up building correlations in the out-moving modes. We explicitly construct an operator which measures this build up of correlation in the frequency space which returns a non-zero expectation value  only 
when the in-state is not a vacuum state. This correlation operator measures the departure of the mode from being in a thermally populated state, which has zero correlation. The diagonal elements of this correlation operator give the spectral profile of the emission of the black hole. We use the modified field correlation spectrum of the  field to extract the initial data. 

We will see that the symmetry characteristics of the initial state will determine whether asymptotic observers can reconstruct the initial state completely or  partially. Therefore, analysis of the allowed set of  symmetries present in the characterization of the initial state will  play a pivotal role in the recovery of the initial information content, just from the spectrum function when the black hole is evaporated completely. Recently, it has been shown \cite{Chatwin-Davies:2015hna} how to reconstruct a qubit state which is thrown into a black hole by measuring changes in the black hole characteristics in such a process. Our scheme is somewhat similar in spirit, but calculates the projection of states in an infinite dimensional Hilbert space. Further, we do not have to rely on measurements of the black hole characteristics prior to and after disturbing it. \textit{We just ask every observer  to report the spectral and correlation profiles they measure, once the black hole has evaporated (practically) 
completely.} The spectra  with a particular kind of distortion, which we call \textit{non-vacuum} distortion, will contain the quantum correlations from which we can  extract the information \cite{Lochan:2015oba} about the initial state. 

In the semi-classical approximation, the formation of the classical event horizon is unavoidable, since the part forming the black hole follows classical equations of motion.  With the presence of the horizon, pure to mixed state transition is also imminent. Thus a part of the field modes always remain hidden from the asymptotic observer, giving rise to a mixed state description. We do not attempt to purify the state through these non-vacuum distortions. Instead, we try to test if the non-vacuum distortions can tell more about the initial quantum data over and above the classical wisdom permitted by no-hair theorems.  We show that, at the semi-classical level itself, there are additional quantum hairs in the resulting radiation profile. At this stage we should emphasize that {\it we are not focusing  on the non-thermal distortions originating from the vacuum itself, but on the part which is originating from non-vacuum component of the quantum state.} As discussed earlier \cite{Visser:2014ypa,Gray:2015pma}, 
there can be various sources of non-thermal corrections, even when the initial  state is a vacuum. We dub the total Hawking radiation endowed with all these corrections, as {\it the vacuum response}. We show that if there are corrections over and above the vacuum response, information regarding the initial sector of Hilbert space, which formed the black hole, can be extracted. 

Information in an initial state of a collapsing field carrying non-zero stress energy can be stored using a superposed state in the Hilbert space. We will be using initial non-vacuum state of the form:
\bea
|\Psi\rangle_{in} &=& \int_0^{\infty} \frac{d \omega}{\sqrt{4 \pi \omega}}f(\omega)\hat{a}^{\dagger}(\omega)|0\rangle_{in}. \label{state} 
\eea
This is an excited state which is a superposition of one-particle states. The function $f(\omega)$ completely encodes the information about the initial state and the idea is to reconstruct this function from the spectrum of the black hole radiation received by asymptotic observers.  

We will first show how this reconstruction of initial state works, for a collapse model forming Schwarzschild black hole in $3+1$ dimension as a quasi-static process.  Thus, we first consider the case in which a spherically symmetric scalar field is undergoing a collapse process to form a black hole. The complete quantum analysis of this process will require the study of the quantum evolution of the field as well as that of the ``quantum geometry" and the back-reaction. However, lack of good control over either the quantum sector of the geometry or the back-reaction in $3+1$ dimension, compels us to adopt a semi-classical approach, where we take the matter field to be described by:
\bea 
\hat{\phi}(x)=\phi_0\mathbb{I} +\delta \hat{\phi},
\eea
where $\phi_0$ is the part which dominantly describes the evolution of geometry. That is, $$\langle \Psi | T_{\mu \nu}[\phi]|\Psi \rangle \sim \langle \Psi | T_{\mu \nu}[\phi_0]|\Psi \rangle$$, will act as the source for the geometry and will lead to the gravitational collapse. We can, alternatively,  think of a process in which some of the highly excited modes $\phi_0$ of the field, acting as classical matter, collapse to form a black hole, while some other modes $\delta \hat{\phi}$ evolve quantum mechanically as a {\it test field} in this background. These {\it quantum} modes are populated, i.e., such modes will be in a non-vacuum state of the field and  carry some small amount of energy into the black hole. Classically, the test field modes, once they cross the horizon, will make the black hole larger and then will become inaccessible to future asymptotic observers. However, quantum mechanically, we show that such a process will lead to a {\it non-vacuum distortion} in the late time Hawking radiation 
which will make the recovery of information about the initial quantum state possible. The late time radiation will have a frequency space correlation which becomes non-zero if the initial state was not a vacuum.

We will also discuss a $1+1$ dimensional  dilatonic black hole solution which is  known as the CGHS model. This model includes  the effect of the back-reaction, in which the problem of forming a black hole with  quantum matter as source can be solved exactly. Therefore, in this model, we will be able to not only account for the back-reaction of the test field modes $\delta \hat{\phi}$ but also recover the information about the field $\phi_0$ which forms the black hole itself. Further, being a conformally flat spacetime, the Bogoliubov  coefficients can be calculated exactly and hence the information content in the out-going modes can be obtained at any stage of evolution, not only at late times. Thus, we can track the loss of information, if any, during the course of evolution and evaporation.

In this paper, we discuss a pure state undergoing a collapse, which forms a black hole. (The case in which the initial data is a mixed state, e.g. a thermal state, will be dealt with separately elsewhere). We identify a special class of initial states whose entire information can be retrieved \textit{just from} the radiation correlation profile of the black hole, without requiring us to analyze other higher correlations of the out-going modes. In \cite{Lochan:2015oba} we showed that if we only focus on the spectrum  (and not on the correlations), the departure from thermality fixes the expectation values of a large set of operators. There is also some special set of initial symmetries which encodes the in-state completely in the distorted spectrum. Expectedly, the correlations carry much more information than the spectrum, are more effective and allows recovery of the complete information about the initial state for a larger class of initial states. 

A more complete treatment of the collapse scenario will involve studying a full quantum gravity analysis. However, since such a theory is still missing, we can introduce another level of sophistication  by introducing  the back-reaction in to the analysis. For the purpose of demonstration, we will be using a dilatonic CGHS black hole in $1+1$ dimension to accommodate the back-reaction. The case of dilatonic black holes can also be promoted gradually towards a full quantum analysis using corrections at various loop orders in the spirit of \cite{Mikovic:1996bh}, which we will pursue elsewhere.

In \ref{Sec_2} and \ref{Sec_02A}, we discuss the characterization of the initial state we will be  dealing with in this paper. We will also briefly discuss quantum field modes in a spherically symmetric black hole collapse scenario. These modes will be used to specify the non-vacuum state of the test field in the spherically symmetric case. In \ref{Sec_03}, we discuss the resulting correlation spectrum from such non-vacuum, single particle states. We will see that the correlation profile stores  information about the initial data.  We also show that the specification of the symmetries of the initial configuration encode more information about the state in the resultant radiation. For some particular states we have retrieved complete information as well. 

After setting up the general framework and demonstrating the concepts in a  semi-classical evaporation model of a Schwarzschild black hole, in which the back-reaction is ignored, we turn to the inclusion of back-reaction in the later sections. Since accommodating the back-reaction in the 3+1 black hole formation is technically very difficult, we go to lower dimensions to get a handle on the  back-reaction analytically. For this purpose we consider the evaporation of a $1+1$ dimensional dilatonic black hole solution. We will briefly describe this model in \ref{Sec06}, and also deal with the semi-classical evaporation in the context of this model, just to connect up with the 3+1 results. In \ref{Sec07}, we will show how the concept of the non-vacuum distortion can be utilized to harness information about the matter falling into the black hole, which normally would have been invisible to the asymptotic (left-moving) observer. We discuss the class of symmetry characteristics of the initial state, for which such 
an asymptotic observer  can reconstruct the initial data. In \ref{Sec08}, we will discuss the implication of our scheme of retrieval of information and the scope for further generalization. 
\section{Initial state of spherical collapse}\label{Sec_2}

The case we discuss first is that of a  real scalar quantum field living in a collapsing spacetime, which eventually would harbor a black hole. The initial state of the field is specified at the past null infinity (${\cal J}^{-}$). The geometry at ${\cal J}^{-}$ is Minkowski-like and the corresponding modes describing the quantum field will be the flat spacetime modes. For the flat spacetime free-field theory, the in-falling field decomposition is given as
\bea
\hat{\phi}({\bf x}) &=& \int \frac{d^3 {\bf k}}{\sqrt{2 \omega_{\bf k}}}(\hat{a}_{{\bf k}}e^{ik \cdot x}+\hat{a}_{{\bf k}}^{\dagger}e^{-ik \cdot x}),\\
                    &=& \int d^3 {\bf k} \hat{\phi}_{\bf k}(t)e^{i{\bf k}\cdot {\bf x}},
\eea
where
\bea
 \hat{\phi}_{\bf k}(t) &=& \frac{(\hat{a}_{{\bf k}}e^{-i\omega_{\bf k} t}+\hat{a}_{{\bf-k}}^{\dagger}e^{i\omega_{\bf k}t})}{\sqrt{2 \omega_{\bf k}}},
\eea
satisfying $\hat{\phi}_{\bf k}=\hat{\phi}_{-{\bf k}}^{*}$ for a real field. We further define $\hat{\bar{\phi}}_{\bf k}=\hat{a}_{{\bf k}}e^{-i\omega_{\bf k} t}$, s.t. 
\bea
\sqrt{2 \omega_{\bf k}} \hat{\phi}_{\bf k} = \hat{\bar{\phi}}_{\bf k} + \hat{\bar{\phi}}^{\dagger}_{\bf -k}.
\eea
Therefore, specifying $\hat{\bar{\phi}}_{\bf k}$ is equivalent to specifying $\hat{\phi}_{\bf k}$. This operator describes the field configuration in terms of the Fourier momentum modes at $\mathcal{J}^{-}$. We define an observable of the momentum correlation by introducing 
the Hermitian operator:
\bea
\hat{N}_{{\bf k_1}{\bf k_2}} &\equiv& \hat{\bar{\phi}}_{\bf k_1} \hat{\bar{\phi}}^{\dagger}_{\bf k_2} + \hat{\bar{\phi}}_{\bf k_2} \hat{\bar{\phi}}^{\dagger}_{\bf k_1} \nonumber\\
&=& \hat{a}_{\bf k_1}\hat{a}^{\dagger}_{\bf k_2}e^{-i(\omega_{\bf k_1}-\omega_{\bf k_2})t}+\hat{a}_{\bf k_2}\hat{a}^{\dagger}_{\bf k_1}e^{i(\omega_{\bf k_1}-\omega_{\bf k_2})t}.  \label{Cor-Op}
\eea
For a massless field in a spherically symmetric configuration, this operator can also measure the frequency correlation if we suppress the angular dependence. We will later concentrate on these cross-correlators in order to retrieve the information about what went into the black hole.
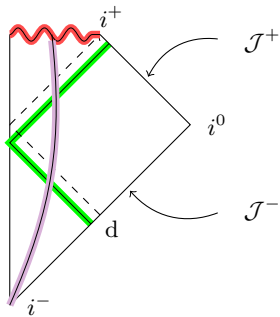
\begin{figure}[h] 
    \begin{center}
        \begin{tikzpicture}[scale=0.8]
        
            \draw[-] (0,-3) -- (0,1.5);
            \draw[dashed,->] (0,0) -- (1.5,1.5);
            \draw[green,line width=1mm] (1.65, 1.35) -- (0,-0.3) -- (1.35,-1.65);
            \draw[black,line width=0.01mm] (1.65, 1.35) -- (0,-0.3) -- (1.35,-1.65);
            \draw[red!70,line width=1mm, snake it] (0,1.5) -- (1.5,1.5);
            \draw[black,line width=0.01mm, snake it] (0,1.5) -- (1.5,1.5);
       
            \draw[black] (0,-3) -- (3,0) -- (1.5,1.5);
            \draw[black, dashed] (1.5,-1.5) -- (0,0);
        
            \draw[violet!30,line width=1mm,bend left=15] (0.7,1.4) to (0,-3);
            \draw[black,line width=0.01mm,bend left=15] (0.7,1.5) to (0,-3);

            \node[label=right:$i^+$] (1) at (1.2,1.8) {};
            \node[label=right:$i^-$] (2) at (0,-3) {};
            \node[label=right:$i^0$] (3) at (3.0,0) {};
            \node[label=below: d](4) at (1.7,-1.3) {};
            
            \node[label=right:$\mathcal{J}^+$] (5) at (3.6,1.4) {};
            \node[label=right:$\mathcal{J}^-$] (6) at (3.6,-1.4) {};

            \draw[->,bend right=35] (5) to (2.3,0.8);
            \draw[->,bend left=35] (6) to (2.0,-1.1);
           
        \end{tikzpicture}
    \end{center}
    \caption{Penrose Diagram For Schwarzschild collapse}
    \label{schwarzschild}
\end{figure}
In this section, we will study a field which undergoes spherically symmetric (s-wave) collapse, slowly \cite{Visser:2014ypa,Gray:2015pma} to form a large mass black hole. The relevant  positive frequency modes describing the initial state will be
\bea
u_{\omega}(t,r,\theta,\phi) \sim \frac{1}{r\sqrt{\omega}}e^{-i\omega(t+r)}S(\theta,\phi)
\eea
where $S(\theta,\phi)$ gives a combination of spherical harmonics $Y_{lm}(\theta,\phi)$. For this collapsing case we take the initial state to be in-moving  at ${\cal J}^{-}$ which is totally spherically symmetric, i.e. $l=0$. Once an event horizon is formed through the collapse process, the full state can again be described using a combined description at the event horizon ${\cal H}$ and on the future null infinity (${\cal J}^{+}$) \cite{Padmanabhan:2009vy,Parker:2009uva}, i.e., the field configuration of spacetime can also be described using positive and negative frequency modes compatible to ${\cal J}^{+}$ as well as on the horizon ${\cal H}$. For an asymptotic observer, the end configuration of the field will be the out-state, described using modes at ${\cal J}^{+}$, which are again flat spacetime modes owing to the asymptotic flatness of the model.  The field content of the out-state can be obtained using the Bogoliubov coefficients between the modes at ${\cal J}^{-}$ and ${\cal J}^{+}$ \cite{
Padmanabhan:2009vy, Parker:2009uva}. The asymptotic form of these Bogoliubov coefficients are given as \cite{Hawking:1974sw}
\begin{widetext}
\bea 
\alpha_{\Omega \omega} &=&\frac{1}{2\pi \kappa} \sqrt{\frac{\Omega}{\omega}}\exp{\left[\frac{\pi\Omega}{2\kappa}\right]}\exp{[i(\Omega-\omega)d]}\exp{\left[\frac{i\Omega }{\kappa} \log{\frac{\omega}{C}} \right]}\Gamma\left[-\frac{i\Omega }{\kappa}\right],\nonumber\\
\beta_{\Omega \omega} &=&-\frac{1}{2\pi \kappa} \sqrt{\frac{\Omega}{\omega}}\exp{\left[-\frac{\pi\Omega}{2\kappa}\right]}\exp{[i(\Omega+\omega)d]}\exp{\left[\frac{i\Omega }{\kappa} \log{\frac{\omega}{C}} \right]}\Gamma\left[-\frac{i\Omega }{\kappa}\right]. \label{BT}
\eea
\end{widetext}
where $\Omega$ is the frequency of the out-modes at ${\cal J}^{+}$, the parameter $\kappa$ is the surface gravity of the black hole, while $C$ is a product of affine parametrization of incoming and outgoing null rays \cite{Padmanabhan:2009vy,Parker:2009uva} and $d$ is an arbitrary constant marking the last null ray reaching ${\cal J}^{+}$. 

We can set $d=0$ through co-ordinate transformations on ${\cal J}^{-}$. These Bogoliubov coefficients are accurate for large values of $\omega$. At small $\omega$ values, the expressions in \ref{BT} will receive corrections. However, when we are interested in  the late-time radiation at future null infinity (${\cal J}^{+}$), one can show that the dominating spectra will come from those modes which have just narrowly escaped the black hole, i.e. which were scattered just before the formation of the event horizon. Such modes are the ones with high frequencies at the past null infinity. So the calculations done with \ref{BT} will be accurate to the leading order.
We will consider states which are eigenstates of the number operator defined with the help of the in-modes, but are not the energy or momentum eigenstates. Let the state of the field undergoing collapse in a black hole spacetime, be a (superposition of) single particle excitation state
\bea
|\Psi\rangle_{in} = \int_0^{\infty} \frac{d \omega}{\sqrt{4 \pi \omega}}f(\omega)\hat{a}^{\dagger}(\omega)|0\rangle. \label{state1}
\eea
(We will  generalize the analysis for higher excited states in the appendices.)
We define a new function $g(z),$ using a dimensionless variable $z$, which is related to the frequency $\omega$ of mode functions at ${\cal J}^{-}$, as  
\bea
\log{\frac{\tilde{\omega}}{C}} =z \Rightarrow f(C e^z)=g(z),  \label{vartrans}
\eea
to re-write the state as
\bea
|\Psi\rangle_{in} = \int_{-\infty}^{\infty} \frac{d z}{\sqrt{4 \pi }}g(z)\hat{a}^{\dagger}(z)|0\rangle. \label{state2}
\eea 
In order to specify the state we need to specify $f(\omega)$ in \ref{state1} or equivalently $g(z)$ in \ref{state2}. We will see how much of  information about this function can be retrieved from the outgoing modes.
Before proceeding to the black hole emission spectra, we introduce the function 
\bea
F(y)= \int_{-\infty}^{\infty} dz g(z) e^{i y z}, \label{F_FT}
\eea
which will be used to characterize the initial state in \ref{state1} or \ref{state2}. This
is an equally good measure for encoding the information about the state, as it is just a Fourier transform of an $\mathbb{L}^2$ function. It is useful to  construct yet another function 
\bea
\tilde{F}\left(\frac{\Omega}{\kappa}\right) = \exp{\left[\frac{\pi\Omega}{2\kappa}\right]} F\left(\frac{\Omega}{\kappa}\right), \label{Ftilde}
\eea
from \ref{F_FT}. Then one  can obtain the distribution $g(t)$ from \ref{Ftilde} as
\bea
g(z)=\frac{1}{2\pi} \int_{-\infty}^{\infty} d \left(\frac{\Omega}{\kappa}\right) \tilde{F}\left(\frac{\Omega}{\kappa}\right) e^{-i \frac{\Omega}{\kappa}z}e^{-\frac{\pi \Omega}{2\kappa}}. \label{Def_gz}
\eea
We note again  that the one particle state is completely specified once we have complete knowledge of the function $\tilde{F}(\Omega/\kappa)$.  We will discuss the information about $\tilde{F}\left(\Omega/\kappa\right)$  available  in the out-going modes, by analyzing the frequency space correlations. Such correlations turn out to be definitive tools for the recovery of the information. We will first discuss these correlations briefly in the next section and then go on to study a special case of this correlation, the self-correlation, which gives the emission profile of the black hole, and we will show that the emission profile develops a non-vacuum, non-thermal, part capable of storing information.
\section{Information of black hole formation : Correlation function}\label{Sec_02A}

We wish to obtain the information regarding the quantum states falling into the black hole, or more elaborately, regarding the quantum states which formed the black hole itself. For this purpose, we consider an observable which measures the frequency space correlation in the out-going modes. The annihilation operator $\hat{b}_{\Omega}$ associated with the outgoing modes is related to the creation and annihilation operator $\hat{a}_{\omega}$ and $\hat{a}^{\dagger}_{\omega}$ of the ingoing mode as,
\begin{align}
\hat{b}_{\Omega}=\int d\omega \left(\alpha _{\Omega \omega}^{*}\hat{a}_{\omega}-\beta _{\Omega \omega}^{*}\hat{a}_{\omega}^{\dagger}\right)
\label{bgt}
\end{align}
where, $\alpha _{\Omega \omega}$ and $\beta _{\Omega \omega}$ are the Bogoliubov coefficients. Now following \ref{Cor-Op}, the frequency space correlation for the outgoing modes turns out to yield 
\begin{align}
\hat{N}_{\Omega _{1}\Omega _{2}}=\hat{b}^{\dagger}_{\Omega _{1}}\hat{b}_{\Omega _{2}}e^{-i(\Omega _{1}-\Omega _{2})t}+\hat{b}^{\dagger}_{\Omega _{2}}\hat{b}_{\Omega _{1}}e^{i(\Omega _{1}-\Omega _{2})t}
\label{frequ-cor}
\end{align}
Recently, an analogous operator was used in \cite{Akhmedov:2015xwa} for studying the growth of loop corrections in an interacting theory. For the  initial state (in-state) of the field, being vacuum $|0\rangle$ or one with a definite momentum $|{\bf k}\rangle$ (and hence for all the Fock basis states), the expectation value of this correlation operator vanishes. We now consider this quantum correlation of the field in the out-going modes. The correlation operator $\hat{N}_{\Omega_{1}\Omega_{2}}$ defined in \ref{frequ-cor} for the out-going modes is related to those for in-moving modes through Bogoliubov transformations, as presented in \ref{bgt}. If the test field $\delta\hat{\phi}$ starts in the in-vacuum state $|0\rangle_{\text{in}}$, then the expectation value of the frequency correlation becomes,
\begin{align}
{}_{\text{in}}\langle 0 |\hat{N}_{\Omega_{1}\Omega_{2}}|0\rangle_{\text{in}}&=\delta (\Omega_1 - \Omega_2) \times e^{\frac{-\pi(\Omega_1 + \Omega_2)}{2\kappa}}\frac{\sqrt{\Omega_1 \Omega_2}}{4 \pi^2 \kappa^2}
\nonumber
\\
&\times\left\lbrace \Gamma \left[-i \frac{\Omega_1}{\kappa} \right] \Gamma \left[i \frac{\Omega_2}{\kappa}\right] e^{-i(\Omega_1 - \Omega_2)t} + c.c.\right\rbrace 
\end{align}
which vanishes identically for the off-diagonal elements and hence {\it the asymptotic future observer will also measure no frequency correlation} in the out-going modes. The diagonal elements of this observable gives the number spectrum. Such an observer measures the out-going spectrum to be a thermal one which can be verified by taking $\Omega_1 = \Omega_2$.

 However, when the test field starts in a non-vacuum state, the out-going modes will develop a frequency correlation and the expectation of \ref{frequ-cor} will become non-zero. The correction to the expectation of the frequency correlator $\hat{N}_{\Omega _{1}\Omega _{2}}$ in a non-vacuum in-state leads to,
\begin{align}
&{}_{\text{in}}\langle \psi |\hat{N}_{\Omega_{1}\Omega_{2}}|\psi\rangle_{\text{in}}=\Bigg[\left(\int \frac{d\omega}{\sqrt{4\pi\omega}}f(\omega)\alpha ^{*}_{\Omega _{2}\omega}\right)
\nonumber
\\
&\times\left(\int \frac{d\bar{\omega}}{\sqrt{4\pi\bar{\omega}}}f^{*}(\bar{\omega})\alpha _{\Omega _{1}\bar{\omega}}\right)
+\left(\int \frac{d\omega}{\sqrt{4\pi\omega}}f^{*}(\omega)\beta ^{*}_{\Omega _{2}\omega}\right)
\nonumber
\\
&\times \left(\int \frac{d\bar{\omega}}{\sqrt{4\pi\bar{\omega}}}f(\bar{\omega})\beta _{\Omega _{1}\bar{\omega}}\right)\Bigg]e^{-i(\Omega _{1}-\Omega _{2})t}
+\textrm{c.c}
\end{align}
Using the expressions for $\alpha _{\Omega \omega}$ and $\beta _{\Omega \omega}$ from \ref{BT}, we obtain,
\begin{align}
{}_{\text{in}}\langle \psi |\hat{N}_{\Omega_{1}\Omega_{2}}|\psi\rangle_{\text{in}}&=\frac{1}{4\pi}\left[A(\Omega_1)A(\Omega_2)^* +c.c.\right]
\nonumber
\\
&+\frac{1}{4\pi}\left[B(\Omega _{1})B(\Omega _{2})^{*}+c.c.\right]. \label{FrqCorr}
\end{align}
with
\bea
A(\Omega)= e^{-\frac{\pi \Omega}{2\kappa}}\frac{\sqrt{\Omega}}{2 \pi \kappa}\Gamma \left[-i \frac{\Omega}{\kappa} \right] F\left(\frac{\Omega}{\kappa}\right)e^{-i\Omega t}, 
\eea
and 
\begin{align}
B(\Omega)=e^{\frac{\pi \Omega}{2\kappa}}\frac{\sqrt{\Omega}}{2\pi \kappa}\Gamma\left[-i\frac{\Omega}{\kappa}\right]F^{*}\left(-\frac{\Omega}{\kappa}\right)e^{-i\Omega t}
\end{align}
expressed in terms of the time co-ordinate of out-going observers. The frequency correlation for two distinctly separated frequencies (i.e., $\Omega_1\neq\Omega_2)$, as discussed above, remains zero for all the field configurations which were in the vacuum (incidentally, also for configurations in individual Fock basis elements) in the in-state. However, as for the out-state, the frequency correlation remains zero {\it only if} the in-state was a vacuum. The out-going modes develop frequency correlation, even if the in-state was a non-vacuum Fock basis state with zero correlation. Alternatively, those fields which carried some amount of stress energy into the black hole definitely develop some non-zero correlation at late times, while only those fields which were in a vacuum state develop no late time frequency correlation. Therefore, by just measuring this operator, a late time observer will be able to tell if some non-zero stress-energy has entered the black hole. 

Further the observer can decipher the state that entered into the black hole, by reconstructing $F(\Omega/\kappa)$ from this non-zero expectation value of the correlation. We demonstrate the technique below. We can also consider a special case  of this correlation, i.e., the self correlation for simplicity. We measure the change in self-correlation, which is just the spectrum operator \cite{Lochan:2015oba}, once a non-vacuum state perturbs the black hole configuration and we will see that this change encodes information of interest (see the appendices). Expectedly, other correlations carry information about the in-going states much more efficiently. Exploration of other correlation functions in this regard will be reported in a subsequent work.
\section{Radiation from Black Hole: Information about the initial state}\label{Sec_03}

We show that \ref{FrqCorr} can be used by out-moving observers for retrieving information regarding $F(\Omega)$ and thus for reconstructing the state presented in \ref{state}. From the off-diagonal elements of \ref{FrqCorr} we construct a complex quantity
\bea
{\cal D}_{\Omega_1 \Omega_2}\equiv N_{\Omega_1 \Omega_2}+\frac{i}{\Delta \Omega}\frac{\partial}{\partial t}N_{\Omega_1 \Omega_2},
\eea
where $\Delta \Omega = \Omega_1 - \Omega_2$. In terms of the function $F(\Omega/\kappa)$, the above expression can be re-written as
\begin{widetext}
\begin{align}
{\cal D}_{\Omega_1 \Omega_2}&=\frac{1}{2\pi}\frac{\sqrt{\Omega_1 \Omega_2}}{4 \pi^2 \kappa^2} 
\Gamma \left[-i \frac{\Omega_1}{\kappa} \right] \Gamma \left[i \frac{\Omega_2}{\kappa}\right] 
\Bigg\lbrace e^{\frac{\pi (\Omega_1 + \Omega_2)}{2\kappa}}  
F\left(-\frac{\Omega_2}{\kappa}\right)F^*\left(-\frac{\Omega_1}{\kappa}\right) 
+e^{-\frac{\pi (\Omega_1 + \Omega_2)}{2\kappa}}  
F\left(\frac{\Omega_1}{\kappa}\right) F^{*}\left(\frac{\Omega_2}{\kappa}\right)\Bigg\rbrace e^{-i(\Omega_1 - \Omega_2)t}. \nonumber\\ \label{InfoExtract1}
\end{align}
\end{widetext}
For a real initial state, we have $F(\Omega/\kappa) = F^*(-\Omega/\kappa)$ and therefore, \ref{InfoExtract1} can be used to extract the function $F(\Omega/\kappa)$ as,
\begin{align}
S_{\Omega_1 \Omega_2}&\equiv \frac{4 \pi^3 \kappa^2 }{\sqrt{\Omega_1 \Omega_2}}\frac{{\cal D}_{\Omega_1 \Omega_2}e^{i(\Omega _{1}-\Omega _{2})t}}{\Gamma \left[-i \frac{\Omega_1}{\kappa} \right] \Gamma \left[i \frac{\Omega_2}{\kappa}\right]\cosh \left(\frac{\pi(\Omega _{1}+\Omega _{2})}{2\kappa}\right) } 
\nonumber
\\
&= F\left(\frac{\Omega_1}{\kappa}\right)F^{*}\left(\frac{\Omega_2}{\kappa}\right). \label{InfoExtract2}
\end{align}
Clearly, left hand side of \ref{InfoExtract2} can be determined by observing the emission spectrum. Therefore, left hand side is under our control completely. From the above relation, we see that this quantity has to be separable as a product in terms of the frequencies $\Omega_1$ and $\Omega_2$. Using this property, we can obtain the function $F(\Omega/\kappa)$, upto an irrelevant constant phase, from the symmetric sum
\bea
\log{S_{\Omega_1 \Omega_2}} = \log{F\left(\frac{\Omega_1}{\kappa}\right)} +\log{F^{*}\left(\frac{\Omega_2}{\kappa}\right)},
\eea
or, alternatively, by fixing one of the frequencies and varying the other. Therefore, for the real initial state, the state can be identically and completely reconstructed from correlations in the out-going modes.

In \cite {Lochan:2014xja}, we devised a formalism to deal with the analysis of field content of a non-vacuum pure states corresponding to a particular observer with respect to another set of observers, using the correlation functions. The information about the state through the function $f(\omega)$, together with the Bogoliubov coefficients, completely characterize the deviations form the standard vacuum response. The analysis of the spectrum operator \cite{Page:1993wv,Birrell:1982ix} also captures this distortion.  The extraction of information about initial data using the spectral distortion $\hat{N}_{\Omega} = \langle \Psi | \hat{b}^{\dagger}_{\Omega}\hat{b}_{\Omega}|\Psi\rangle -\langle 0 | \hat{b}^{\dagger}_{\Omega}\hat{b}_{\Omega}|0 \rangle$, as reported in \cite{Lochan:2015oba} is presented in detail in \ref{App01}. In \ref{App_02} and \ref{App_03}, we show that for a particular class of symmetric initial states, interesting quantities can be obtained from the out-states as well as their 
generalization for multi-particle states. The frequency correlator for a multi-particle state gives the information about the one-particle sector through the reduced density matrix as shown in \ref{AppB}. Higher order correlation functions will give the information regarding the many particle sectors subsequently. However, in this paper, we only focus on the single particle case, for simplicity. 

We have thus demonstrated the existence of semi-classical hairs in the case of the spherically symmetric collapse which would have formed a Schwarzschild black hole classically. We learn that if we are aware of the symmetries of the system which is going to form a black hole, from some general principles,  we will know how the non-vacuum response would look like. We can measure particle content for different test fields. The test field which contributes infinitesimal  energy to the formation will reflect its non-vacuum character in the late time radiation. That is to say, its spectra will show deviations from the expected vacuum response, corresponding to the symmetries of initial data. Measurements of such non-vacuum distortion will reveal partial or complete character of the state of the field depending on the knowledge of the symmetry of initial profile. 

The cases discussed above were all based on test field approximations. We can extrapolate this idea to conjecture that if we have correct account of the back-reaction, or quantum gravity corrected Bogoliubov coefficients, they will still provide a handle for initial data as in \ref{InfoExtract1} (see also \ref{NCEx1} in appendix). It will be worthwhile to demonstrate these ideas for a  set-up including the back-reaction. In a general 3+1 collapse scenario the precise handling of back-reaction remains an open problem. Even for the spherical collapse case, which we discussed above, accounting for the back-reaction is a tedious task. We will instead be looking at a $1+1$ dimensional dilatonic CGHS black hole model. In this case the issue of back-reaction can be exactly handled and even the full quantum gravity calculation can be implemented perturbatively. However, in this paper, we will  be content with the semi-classical scheme wherein the  back-reaction of the test field has been accounted for. We will see 
how the non-vacuum distortions lead to additional quantum hairs, which would have been missed classically. 
\section{CGHS model: Introduction}\label{Sec06}

The CGHS black hole solution \cite{Callan:1992rs,Fabbri:2005mw} is a $1+1$ dimensional gravity model of a dilatonic field  $\phi$ (along with possibly other matter fields). The theory will be described by the action,
\begin{widetext}
\begin{align}
\mathcal{A}=
\frac{1}{2\pi}\int d^{2}x\sqrt{-g}\left[e^{-2\phi}\left(R+4(\mathbf{\nabla} \phi)^{2}+4\lambda ^{2}\right)-\frac{1}{2}\sum _{i=1}^{N}(\mathbf{\nabla}f_{i})^{2} \right]. \label{CGHS_Action}
\end{align}
\end{widetext}
where $\lambda ^{2}$ is the cosmological constant and  $f_{i}$ stands for $i-$ th matter field; $N$ such total fields may be present. Since all two dimensional space-times are conformally flat the metric ansatz will involve a single unknown function, the conformal factor, which is written in double null coordinates as,
\begin{align}
ds^{2}=-e^{2\rho}dx^{+}dx^{-}.
\end{align}
For the matter fields, the classical solutions are those in which, $f_{i}(x^{+},x^{-})=f_{i+}(x^{+})+f_{i-}(x^{-})$. Then, given some particular matter fields, one can obtain corresponding solutions for $\phi$ and $\rho$ respectively. The simplest among all of them corresponds to the vacuum solution in which $e^{-2\rho}=e^{-2\phi}=(M/\lambda)-\lambda ^{2}x^{+}x^{-}$. This represents a black hole of mass $M$, with line element,
\begin{align}
ds^{2}=-\frac{dx^{+}dx^{-}}{\frac{M}{\lambda}-\lambda^2 x^{+}x^{-}},
\end{align}
while in absence of any mass, i.e., $M=0$  we obtain the linear dilatonic vacuum solutions as
\begin{align}
ds^{2}=-\frac{dx^{+}dx^{-}}{-\lambda^2 x^{+}x^{-}}. \label{DilatonVacuum}
\end{align}
A more realistic and dynamical situation corresponds to the case when an in-coming matter forms a singularity. If the matter starts at $x_{i}^{+}$ and extends up to $x_{f}^{+}$, then the line element  turns out to be,
\begin{align}
ds^{2}=-\frac{dx^{+}dx^{-}}{\frac{M(x^{+})}{\lambda}-\lambda ^{2}x^{+}x^{-}-P^{+}(x^{+})x^{+}},
\end{align}
where  $M(x^{+})$ and $P^{+}(x^{+})$ correspond to the integrals,
\bea
M(x^{+})&=&\int _{x_{i}^{+}}^{x^{+}}dy^{+}y^{+}T_{++}(y^{+}), \label{MassInt}\\
P^{+}(x^{+})&=&\int _{x_{i}^{+}}^{x^{+}}dy^{+}T_{++}(y^{+}).
\eea
The region outside $x_{f}^{+}$ is a black hole of mass $M\equiv M(x_{f}^{+})$. One can check that there is a curvature singularity where the conformal factor diverges. 

The singularity hides behind an event horizon for future null observers receiving the out-moving radiation. The location of the event horizon can be obtained starting from the location of the apparent horizon. This can, in turn,  be obtained using $\partial _{+}A\leq 0$, where $A$ stands for the transverse area. Using the four dimensional analog we end up getting, $\partial _{+}e^{-2\phi}\leq0$. Here the equality would lead to the location of the event horizon beyond $x^{+}_{f}$, which happens to be at $x^{-}_{h}=-P^{+}/\lambda ^{2}$.
\begin{figure}[h]
    \begin{center}
        \begin{tikzpicture}[scale=0.8]
     
            \draw[dashed,->] (-1.5,-1.5) -- (0,0) -- (1.5,1.5);
        
            \draw[orange, snake it] (-1.5,1.5) -- (1.5,1.5);
              \draw[black] (0,-3) -- (3,0) -- (1.5,1.5);
            \draw[black] (0,-3) -- (-3,0) -- (-1.5,1.5);
            \draw[blue] (1.5,-1.5) -- (0,0) -- (-1.5,1.5); 
             \draw[blue] (1.60,-1.40) -- (-1.3,1.50);
             
             \draw[green, thick] (-1.4,-1.6) -- (1.6,1.4);
             
            \draw[black, dotted, thick, bend left=15] (0,-3) to (-1.5,1.5);
            \draw[magenta, bend right=15] (0,-3) to (1.5,1.5);


            \node[label=right:$\mathcal{J}_R^+$] (4) at (3.6,1.4) {};
            \node[label=right:$\mathcal{J}_R^-$] (5) at (3.6,-1.4) {};
            \node[label=below: $x_i^+$](6) at (1.5,-1.4) {};
            \node[label=below: $x^-$](6) at (-0.75,-2.2) {};
            \node[label=left:$\mathcal{J}_L^+$] (7) at (-3.6,1.7) {};
            \node[label=left:$\mathcal{J}_L^-$] (8) at (-3.6,-1.4) {};
             \node[label=below: $x_f^+$](9) at (2.1,-0.9) {};
             \node[label=below: $x^+$](10) at (0.75,-2.2) {};
           \node[label=above: $i_L^{+}$] at (-1.5,1.5) {};
            \node[label=above: $i_R^{+}$] at (1.5,1.5) {};
             \node[label=below: $i^{-}$] at (0,-3) {};

            \draw[->,bend right=35] (4) to (2.3,0.8);
            \draw[->,bend left=35] (5) to (2.5,-1.0);
            \draw[->,bend left=35] (7) to (-2.3,0.8);
            \draw[->,bend right=35] (8) to (-2.5,-0.8);

        \end{tikzpicture}
    \end{center}
    \caption{A CGHS black hole}
    \label{fig:penrose}
\end{figure}
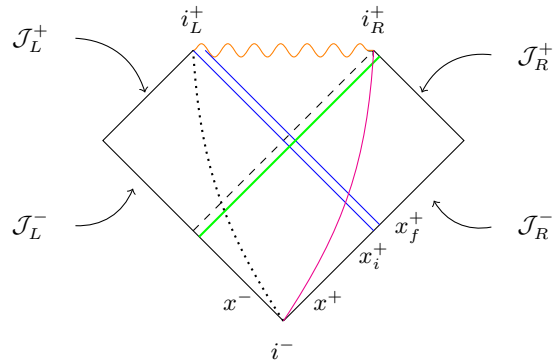
Thermodynamics, as well as Hawking evaporation of such a black hole solution (also with different matter couplings), have been extensively studied \cite{Fabbri:2005mw,Birnir:1992by,Giddings:1992ff,Alves:1995hn,Alves:1999ct}. Even higher loop corrections to the thermal character of vacuum response have been studied \cite{Mikovic:1996bh}. We will demonstrate our ideas for the simplest coupling as shown in \ref{CGHS_Action}. 

We first study the genesis of Hawking radiation in such a black hole formation. The field equations for this action, at the classical level are written as
\bea
-\partial_+\partial_-e^{-2\phi}-\lambda^2 e^{2 \rho-2\phi} = 0,  \label{CGHS-EOM1}\\
2e^{-2\phi}\partial_+\partial_-(\rho -\phi)+\partial_+\partial_-e^{-2\phi}+\lambda^2 e^{2 \rho-2\phi} = 0, \label{CGHS-EOM2}
\eea
since the stress energy tensor decouples into left-moving and right-moving parts.
Further, the two constraint equations  involve the energy momentum tensor components on both the initial slices and lead to,
\begin{subequations}
\begin{align}
\partial _{+}^{2}e^{-2\phi}+4\partial _{+}\phi \partial _{+}\left(\rho -\phi \right)e^{-2\phi}+T_{++}=0,
\\
\partial _{-}^{2}e^{-2\phi}+4\partial _{-}\phi \partial _{-}\left(\rho -\phi \right)e^{-2\phi}+T_{--}=0.  \label{CGHS-constraint}
\end{align}
\end{subequations}
Solving the equations of motion \ref{CGHS-EOM1}, \ref{CGHS-EOM2}, we obtain the conformal gauge $ \rho = \phi,$ where the constraint equations are expressed as
\begin{subequations}
\begin{align}
\partial _{+}^{2}e^{-2\phi}+ T_{++} = 0,
\\
\partial _{-}^{2}e^{-2\phi}+ T_{--} = 0.
\end{align}
\end{subequations}
The solution of these equations provide the classical geometry which is depicted in \ref{fig:penrose}. The spacetime prior to $x_i^+$ is flat while the
spacetime beyond $x_f^+$ is described by the black hole geometry. 

For carrying out the semi-classical analysis on this spacetime, we introduce co-ordinate systems suited for $\mathcal{J}_{L}^{-}$ and $\mathcal{J}_{R}^{+}$ respectively. We have the co-ordinate set $z^{\pm}$
\bea
\pm \lambda x^{\pm} = e^{\pm \lambda z^{\pm}},
\eea
which maps the entire $\mathcal{J}_{L}^{-}$ into $z^{-}\in(-\infty,\infty)$. We obtain another co-ordinate system suited for $\mathcal{J}_{R}^{+}$ as
$\sigma _{\rm out}^{\pm}$. Let us first discuss the transformation between $(z^{+},z^{-})$ and $(\sigma _{\rm out}^{+},\sigma _{\rm out}^{-})$  defined as
\begin{align}
z^{+}=\sigma _{\rm out}^{+};\qquad z^{-}=-\frac{1}{\lambda}\ln \left[e^{-\lambda \sigma _{\rm out}^{-}}+\frac{P^{+}}{\lambda} \right].
\end{align}
The horizon, located at $x^{-}=-P^{+}/\lambda ^{2}$, will get mapped to $z^{-}=z_i^{-}= -\frac{1}{\lambda}\log{(P^{+}/\lambda)}$ in these co-ordinates. `In' state modes are defined on the asymptotically flat region $\mathcal{J}_{L}^{-}$ moving towards $\mathcal{J}_{R}^{+}$ and the convenient basis modes can be taken to be,
\begin{align}
u_{\omega}=\frac{1}{\sqrt{2\omega}}e^{-i\omega z^{-}},
\end{align}
where $\omega >0$. The `out' region corresponds to $\mathcal{J}_{R}^{+}$ which receives the state from $\mathcal{J}_{L}^{-}$ after the black hole has formed. The basis modes in the out region at $\mathcal{J}_{R}^{+}$ are
\begin{align}
v_{\omega}=\frac{1}{\sqrt{2\omega}}e^{-i\omega \sigma _{\rm out}^{-}}\Theta (z_i^{-}-z^{-}),
\end{align}
where $\Theta$ is the usual step function arising from the fact that the out modes are supported by states on $\mathcal{J}_{L}^{-}$ only in the interval $(-\infty,0)$. Again the field can be specified fully on $\mathcal{J}_{L}^{-}$ or jointly on $\mathcal{J}_{R}^{+}$ and on the event horizon ${\cal H}_R$. Since the mode functions at ${\cal H}_R$ correspond to part of the field falling into the singularity, which cannot be detected by observers at $\mathcal{J}_{R}^{+}$, they need to be traced over. Thus, the precise form of mode decomposition on ${\cal H}_R$ does not affect physical results for $\mathcal{J}_{L}^{-}$. Therefore, we can expand the dilaton field in a different mode basis as,
\begin{align}
f&=\int _{0}^{\infty}d\omega ~\left[a_{\omega}u_{\omega}+a_{\omega}^{\dagger}u_{\omega}^{*}\right],~~~(\textrm{in})
\\
&=\int _{0}^{\infty}d\omega ~\left[b_{\omega}v_{\omega}+b_{\omega}^{\dagger}v_{\omega}^{*}+\hat{b}_{\omega}\hat{v}_{\omega}+\hat{b}_{\omega}^{\dagger}\hat{v}_{\omega}^{*}\right],~~~(\textrm{out})
\end{align}
where $a_{\omega}^{\dagger}$ corresponds to creation operator appropriate for the `in' region. Similarly $b_{\omega}^{\dagger}$ and $\hat{b}_{\omega}^{\dagger}$ stand for the creation operators for the `out' region and the black hole interior region respectively. The inner product between $v_{\Omega}$ and $u_{\omega}^{*}$ corresponds to,
\begin{widetext}
\begin{align}
\alpha _{\Omega \omega}&=-\frac{i}{\pi}\int _{-\infty}^{z_i^{-}}dz^{-}v_{\Omega}\partial _{-}u_{\omega}^{*}
=\frac{1}{2\pi}\sqrt{\frac{\omega}{\Omega}}\int _{-\infty}^{z_i^{-}}dz^{-}\exp \left[\frac{i\Omega}{\lambda}\ln \left\lbrace \left(e^{-\lambda z^{-}}-\frac{P^{+}}{\lambda}\right) \right\rbrace+i\omega z^{-} \right]
\nonumber
\\
&=\frac{1}{2\pi \lambda}\sqrt{\frac{\omega}{\Omega}}\left(\frac{P^{+}}{\lambda}\right)^{i(\Omega-\omega)/\lambda}~B\left(-\frac{i\Omega}{\lambda}+\frac{i\omega}{\lambda},1+\frac{i\Omega}{\lambda} \right),\label{AlphaCGHS}
\end{align}
while the inner product between  $v_{\Omega}$ and $u_{\omega}$ gives
\begin{align}
\beta _{\Omega \omega}&=\frac{i}{\pi}\int _{-\infty}^{z_i^{-}}dz^{-}v_{\Omega}\partial _{-}u_{\omega}
=\frac{1}{2\pi}\sqrt{\frac{\omega}{\Omega}}\int _{-\infty}^{z_i^{-}}dz^{-}\exp \left[\frac{i\Omega}{\lambda}\ln \left\lbrace \left(e^{-\lambda z^{-}}-\frac{P^{+}}{\lambda}\right) \right\rbrace-i\omega z^{-} \right]
\nonumber
\\
&=\frac{1}{2\pi \lambda}\sqrt{\frac{\omega}{\Omega}}\left(\frac{P^{+}}{\lambda}\right)^{i(\Omega+\omega)/\lambda}~B\left(-\frac{i\Omega}{\lambda}-\frac{i\omega}{\lambda},1+\frac{i\Omega}{\lambda} \right), \label{BetaCGHS}
\end{align}
\end{widetext}
where $B(x,y)$ is the Beta function.
The vacuum response can be obtained from \ref{BetaCGHS}, which has a thermal profile in the limit of large frequencies, which will be the late time limit for observers at $\mathcal{J}_{R}^{+}$. In general the vacuum response also comes with a non-thermal part.  This late-time thermal response originates from the vacuum state at  $\mathcal{J}_{L}^{-}$. But, due to asymmetry in the left- and right-moving modes, the modes which form the black hole, originate only from  $\mathcal{J}_{R}^{-}$, rather than from  $\mathcal{J}_{L}^{-}$. If one has to retrieve the information regarding the quantum states that formed the black hole in the first place, the observers moving left are the relevant ones. Therefore, we will concentrate on the future, left moving, observers who are moving in the flat spacetime throughout. We want to study the black hole evaporation at the semi-classical level for one such observer, which we will do in the next section.
\section{Information regarding the collapsing matter}\label{Sec07}

The left-moving matter is introduced for $x^+ \geq x_i^+$ and hence the spacetime is in a vacuum configuration prior to it. The spacetime, in the region $x^+ < x_i^+$ is flat and the metric is given as in the \ref{DilatonVacuum}, which, on using the co-ordinate
\begin{align}
x^{+}=-\frac{1}{\lambda y^{+}};\qquad x^{-}=-\frac{1}{\lambda y^{-}},\label{coordinateSet01}
\end{align}
remains the same 
\bea
ds^2 =- \frac{d y^+ d y^-}{-\lambda^2 y^{+}y^{-}}.
\eea
The location $x_i^+$ is marked by $y_i^+ =-1/\lambda x_i^+$. Using another set of co-ordinate transformations the metric on $\mathcal{J}_{R}^{-}$ can be brought into the flat form. A left-moving observer who completely stays in the region in the past of ${\cal J}_L^+$ remains in flat spacetime. 

However, such an observer is able to access only portion of initial data on $\mathcal{J}_{R}^{-}$. Therefore, the Bogoliubov coefficients between a complete set of mode functions $u_{\omega}$ defined on  $\mathcal{J}_{R}^{-}$ and the complete set of mode functions $v_{\omega}$ defined on  $\mathcal{J}_{L}^{+}$, are given as
\begin{widetext}
\begin{align}
\alpha _{\Omega \omega}&=-\frac{i}{\pi}\int _{-\infty}^{\chi_i^+}d\chi^+v_{\Omega}\partial _{-}u_{\omega}^{*}
=\frac{1}{2\pi}\sqrt{\frac{\omega}{\Omega}}\int _{-\infty}^{\chi_i^+}d\chi^+\exp \left[\frac{i\Omega}{\lambda}\ln \left\lbrace\left(e^{-\lambda \chi^+}-|y_i^+| \right) \right\rbrace+i\omega \chi^+ \right]
\nonumber
\\
&=\frac{1}{2\pi \lambda}\sqrt{\frac{\omega}{\Omega}}|y_i^+|^{\frac{i(\Omega-\omega)}{\lambda}}~B\left(-\frac{i\Omega}{\lambda}+\frac{i\omega}{\lambda},1+\frac{i\Omega}{\lambda} \right),\label{AlphaCGHS-R-L}
\end{align}
and
\begin{align}
\beta _{\Omega \omega}&=\frac{i}{\pi}\int _{-\infty}^{\chi_i^+}d\chi^+v_{\Omega}\partial _{+}u_{\omega}
=\frac{1}{2\pi}\sqrt{\frac{\omega}{\Omega}}\int _{-\infty}^{\chi_i^+}d\chi^+\exp \left[\frac{i\Omega}{\lambda}\ln \left\lbrace  \left(e^{-\lambda \chi^+}-|y_i^+|\right) \right\rbrace-i\omega \chi^+ \right]
\nonumber
\\
&=\frac{1}{2\pi \lambda}\sqrt{\frac{\omega}{\Omega}}|y_i^+|^{\frac{i(\Omega+\omega)}{\lambda}}~B\left(-\frac{i\Omega}{\lambda}-\frac{i\omega}{\lambda},1+\frac{i\Omega}{\lambda} \right). \label{BetaCGHS-R-L}
\end{align}
\end{widetext}
Thus, as before, the observer at $\mathcal{J}_{L}^{+}$ will observe  Bogoliubov coefficients similar to the ones observed by their right counterparts but with the parameter exchange $|y_i^+| \leftrightarrow P^{+}/\lambda$ \cite{Lochan:2016cxt}. However, these nontrivial Bogoliubov coefficients are totally due to the tracing over of the modes which lie in the future of ${\cal J}_L^+$ and not due to any geometry change. One can check that if the fraction of the tracing over vanishes, which corresponds to the limit $|y_i^+|\rightarrow 0$, the Bogoliubov coefficients will assume a trivial form. Along identical lines, the Bogoliubov coefficients in \ref{AlphaCGHS} and \ref{BetaCGHS} assume a trivial form in the limit $P^+ \rightarrow 0$. For these observers the effect of tracing over is indistinguishable from that of geometry change. Both these effects vanish simultaneously in the above limit. However, the vacuum response for both these observes is indistinguishable.  Late time radiation for such observers on $\
mathcal{J}_{L}^{+}$  for the vacuum state (of a test field) on $\mathcal{J}_{R}^{-}$  is also thermal {\it with the same temperature as measured by their right-moving counterparts !} 

Further, any non vacuum state on $\mathcal{J}_{R}^{-}$ will lead to non-vacuum distortions in the radiation. We will now use the Bogoliubov coefficients for  extracting information regarding the matter that formed the black hole.
\subsection{Test Field approximation}

To start with, we can first do a quick demonstration, under the test field approximation,  in order to connect up with the earlier case of the Schwarzschild black hole, by assuming that the matter $\phi$ forming the black hole is classical. Then we add a little more matter  $\delta \hat{\phi}$ to the black hole perturbatively. That is to say, we add another small matter pulse to the collapse with  support in the region $x^+ > x_f^+$. This small chunk is to be treated perturbatively as quantum matter. Given the form of the Bogoliubov coefficients as in \ref{AlphaCGHS-R-L} and \ref{BetaCGHS-R-L}, we note that they remain independent of the matter content in the region $x^+ > x_i^+$ and solely depend on the co-ordinate at which the first matter shell was introduced, which serves as a horizon for the left-moving observers. Therefore, an asymptotically left-moving observer will use \ref{AlphaCGHS-R-L} and \ref{BetaCGHS-R-L} to compute the spectral distortion and reconstruct the state of the test matter which is 
thrown in, through a procedure similar to what was done for the Schwarzschild black hole. Then, for asymptotic observers at late times,  the high frequency approximation of  \ref{AlphaCGHS-R-L} and \ref{BetaCGHS-R-L} leads to a form exactly as in \ref{BT}. Thus, for such observers, the results developed in \ref{Sec_03}, \ref{App_02} and \ref{App_03} are exactly applicable.
\subsection{Adding back-reaction: Extracting information about the matter forming the black hole}

In the CGHS case, we can do away with the high frequency approximation to exactly calculate the results for observers at any finite time (not necessarily at late times) as well. Moreover,  the results derived for the left-moving asymptotic observers remains oblivious to the  matter introduced beyond their perceived horizon and hence does not care about the geometry beyond the horizon as well. Therefore, all the back-reaction of the collapsing matter can now be accounted for, since they {\it do not} change the geometry profile of left-moving observers {\it at all}. Being a conformally flat spacetime, we  know the exact mode functions irrespective of the back-reaction in the entire spacetime as well as in the relevant left portion. Therefore, the form of \ref{AlphaCGHS-R-L} and \ref{BetaCGHS-R-L} are exact, even in the presence of back-reaction of the  field, or even when we take the collapsing matter $\phi$ itself to be quantum matter, which we will do next.

For a complete semi-classical treatment, we take the field $\hat{\phi}$ to be quantum mechanical. At the semi-classical level, the stress energy tensor components are replaced by their expectation values $\langle T_{\pm \pm}\rangle$ and $\langle T_{+-} \rangle$. Being a two dimensional spacetime the expectation values get an additional contribution  from the conformal anomaly. Therefore, the classical equations (for $N=1$) are modified to 
\bea
-\partial_+\partial_-e^{-2\phi}-\lambda^2 e^{2 \rho-2\phi} = \frac{\hbar}{12\pi}\partial_+\partial_-\rho,  \hspace{0.3 in}  \label{CGHS-SC-EOM1}\\
2e^{-2\phi}\partial_+\partial_-(\rho -\phi)+\partial_+\partial_-e^{-2\phi}+\lambda^2 e^{2 \rho-2\phi} = 0, \label{CGHS-SC-EOM2}
\eea
whereas the constraint equations also pick up conformal anomaly corrections as
\begin{subequations}
\begin{align}
\partial _{+}^{2}e^{-2\phi}+4\partial _{+}\phi \partial _{+}\left(\rho -\phi \right)e^{-2\phi}+\langle T_{++}\rangle=0,
\\
\partial _{-}^{2}e^{-2\phi}+4\partial _{-}\phi \partial _{-}\left(\rho -\phi \right)e^{-2\phi}+\langle T_{--}\rangle=0. \label{SC-Constraints}
\end{align}
\end{subequations}
In order to remain true to the classical geometry, the state of the matter field should be one in which the classical flat geometry is realized prior to $x_i^+$. Thus, we require $\exp(\rho)=\exp(\rho_{\textrm{flat}}) =1/\lambda^2x^+x^-$ in that region suggesting that the matter support is only in the region $x^+ >x_i^+ $. Therefore, the classical values of the $T_{\pm \pm}$ are realized by $ \langle T_{\pm \pm} \rangle$. We can also judiciously choose the boundary conditions for the set of initial states such that the contribution due to conformal anomaly can be canceled in the region of interest, giving rise to a flat spacetime semi-classically, see for instance \cite{Vaz:1996kh, Ashtekar:2010hx, Hawking:1992cc}. We define ${\cal I}_L^+$ as the line $x^+ = x_i^+$. For our consideration, we will need the part of the spacetime in the causal past of ${\cal I}_L^+$, which remains unaffected by the conformal anomaly with such a judicious choice of family of quantum states. 

Supported by such quantum states, the geometry  of the spacetime remains as discussed above and we can use the expressions for the Bogoliubov coefficients as earlier. The asymptotic expressions for the Bogoliubov coefficients as in \ref{AlphaCGHS-R-L} and \ref{BetaCGHS-R-L} resemble those of the spherical collapse model \ref{BT}. Therefore, the spectral distortion for the late time observers will be exactly as discussed in \ref{Sec_03}, \ref{App_02} and \ref{App_03}.

We can now obtain the exact expression and the symmetry profile of the initial data required for the information retrieval by a generic observer on ${\mathcal J}_L^{+}$. Using the Bogoliubov coefficients in \ref{AlphaCGHS-R-L} and \ref{BetaCGHS-R-L} we can obtain the non-vacuum correction to the correlator and the vacuum spectrum through \ref{Cor-Op} (see also \ref{NCEx1}). Again, we will first consider the case of a single particle state as in \ref{state}, which has the stress energy support as discussed above  (The discussion of the multi-particle state will follow along the lines as demonstrated for the Schwarzschild case in \ref{MultiParticle}, also refer to \ref{Familyofstates}). The symmetry profile required in the initial state, for the retrieval of information about the initial state by the late time observers, remains exactly the same as that for the spherical symmetric Schwarzschild model. Therefore, such symmetry profiles appear uniquely for all late time observers and {\it any real initial data 
can be uniquely reconstructed by the late time observers}. 

However, since we have exact expressions for the Bogoliubov coefficients, we can also obtain symmetry condition for all observers on ${\mathcal J}_L^{+}$ and not only for the late time observers. This will demonstrate the ability to reconstruct the information at any stage of evolution, which can be used to track the information content throughout the evolution and quantify whether there is any information loss for a more general initial data. We define another function $\tilde{g}(\omega)$, as
\bea
f(\bar{\omega}') = \bar{\omega}' \tilde{g}(\bar{\omega}') e^{-\frac{\pi}{2}\bar{\omega}'}\Gamma[-i\bar{\omega}']\left|y_i^{+}\right|^{-i\bar{\omega}'}. \label{Trns1}
\eea 
The transformation presented in \ref{Trns1} relates the correlation to the symmetries of $\tilde{g}(\bar{\omega})$. The scheme of information retrieval from the frequency correlator can be implemented exactly as before, for $\tilde{g}(\bar{\omega})$. Reconstruction of $\tilde{g}(\bar{\omega})$ is equivalent to reconstruction of $f(\bar{\omega})$. (Analysis for the self-correlator, i.e., the spectrum operator is discussed in \ref{App_H}.) Different set of observers require different set of symmetries in order to extract maximum amount of information from the non-vacuum correlator. For a given initial state we can also follow the information content during the course of evaporation, which culminates in the late time result.

Thus, using a consistent semi-classical treatment we could recover the matter quantum state using the distortion of the thermal radiation as detected by the asymptotic observers. However any asymptotic observer would have measured a thermal radiation which is independent of the black hole's mass --- and decided entirely by the cosmological constant --- in the model. This   remains true throughout, while the black hole evaporates. For the left moving asymptotic observer, the black hole region does not shrink as an outcome of this radiation emission, since the location of horizon is not decided by the mass content inside the horizon. Still such an observer may be able to reconstruct all the information about the mass/energy content beyond her horizon. Therefore, it is reasonable to expect that such observers associate a notion of entropy to the black hole which is significantly different from what a right moving observer will do. Hence it is expected that the notion of entropy for left-moving observers should 
also substantially differ from the standard black hole entropy expression. In the Schwarzschild black hole formation on the other hand, due to spherical symmetry, left-moving or right moving observers not only measure the same temperature, but witness an identical geometry change during the formation or  evaporation of the black hole. Therefore, unlike the current case, they should be associating the same entropy expression for the hole.  We will pursue these topics elsewhere.
\section{Conclusions}\label{Sec08}

The mystery of loss of information in the black hole physics has two primary aspects. The first, and possibly more troubling aspect, has to deal with the character of the Hawking radiation which, due to presence of the event horizon, is obliged to have a mixed state description. Since Hawking radiation extracts energy out of a black hole, the real problem manifests itself when the black hole gets completely evaporated by this process. The resulting radiation profile remains mixed and hence does not respect the unitary evolution scheme in which a pure state should not have evolved into the mixed state.

A possibly related version of the  information loss problem is due to the classical no-hair conjecture. Classically, whatever ends up inside the event horizon is `visible' to the observers in the exterior region only through certain classical charges. All other information defining the initial data gets destroyed once the matter hits the singularity. That is true as regards the matter forming the black hole  as well.

In this paper, we have focused on the reconstruction of initial data which formed the black hole by observing a particular kind of distortion to the Hawking radiation. For some matter to end up inside the horizon (carrying some energy, charge etc. with it), the state of the matter field should be non-vacuum. Moreover, the non-vacuum nature of the state manifests itself in the correlation between the modes escaping the capture by the black hole and the modes entering the horizon and eventually hitting the singularity. Observing the portion which escaped the horizon, we can re-construct the correlation and hence the state of the field at the initial slice.  We construct an observable, which can capture this correlation. Using this observable judiciously, one can 
extract information pertaining to the initial condition, which remains otherwise hidden from the asymptotic observers. The diagonal elements of this correlation matrix give the distortion of the emission spectrum, which becomes non-thermal for non-vacuum in-states. It is noteworthy that the non-vacuum distortions will always be present in the most general case including quantum gravity, back-reaction etc. Only the corresponding Bogoliubov coefficients will be generalized for the particular case under consideration.

In 3+1 dimensional spacetime, the  inclusion of semiclassical back-reaction is not under control. For the purpose of demonstration of our ideas we first discussed a semi-classical case of a spherical collapse which forms a black hole through a slow, s-wave process. On top of this evolving geometry we introduced another test field, described by a massless scalar field, initially set in a particular field configuration at asymptotic, past, null infinity. The complete recovery of the initial data corresponds to the deciphering of the quantum state of the field  unambiguously.
For this simplified set up, the vacuum response is thermal, and the non-vacuum response corresponds to non-thermal distortions of the Hawking radiation. Such distortions constrain some operators acting on the conjugate space of the frequency representation of the quantum field. Typically such operators provide information about the single particle sectors of the Hilbert space. Since the spectral distortion is just a single function of the frequency, it is expected to capture the finite correlations of an arbitrary, general, initial state. Higher order correlations may be required to obtain additional information about the initial state. 

If (i) the state of the field corresponds to a single particle state and (ii) we have access to the symmetries of the distribution in the frequency space, we can recover a lot more information about the initial state through such non-vacuum distortions. In particular, we show that the existence of a class of symmetries of non-zero measure  will encode all the  information about the initial state in the non-vacuum distortion. In particular,  for real initial data, we can recover the state of a single excitation, {\it completely}, using frequency correlation in the out-going modes. Although the non-vacuum distortions do not make the  out-state a pure one, there are enough correlations available  even in the mixed state, to reconstruct the initial state. Also we did not need to study higher order correlations to obtain further information about the initial state.

The simple case we discussed first, was without the back-reaction  in a non-vacuum configuration.  More generally,  we need to account for the the back-reaction as well, howsoever small they might be. The Bogoliubov coefficients incorporating the back-reaction  will be obtained through the mode functions of in- and out- configurations in such a modified geometry. In general it is a very difficult job. However, since addressing the issue of back-reaction in some simplified scenarios should  throw some light on the concepts involved, we considered the case of back-reaction in a $1+1$  dimensional dilatonic black hole solution, viz., the CGHS model. A dilatonic vacuum solution is perturbatively unstable towards forming a black hole  if the perturbation is around the in-moving modes. Even at the classical level the exact Hawking radiation of this model is thermal at the high frequency regime with corrections at low energies. We call such a radiation profile collectively as the vacuum response.  We showed that 
the non-vacuum distortions for the left-moving observers reveal information about initial left-moving distribution which collapsed to form the black hole. There is always a part of past null infinities, which is causally disconnected to the asymptotic future observers. Classically no information pertaining to the matter field configuration in this disconnected sector will ever be available to such observers. However, we see that the non-vacuum distortions are capable of revealing the information about such configurations through correlations. The modes appearing as distortions at ${\cal J}^{+}$ did have, in the past, some correlations, with the modes which enter the horizon, at ${\cal J}^{-}$. Measurement of such distortions can, in principle, also tell us about the white hole region if it were accompanying the black hole region as suggested in some papers 
\cite{Stephens:1993an,Ashtekar:2005cj,Haggard:2014rza}. A simple extrapolation from this idea can also be used to study the fate of such information regarding the matter field which formed the black hole. The retrieval process for the late time observers is {\it exactly the same} as the Schwarzschild late time observers, yielding the same result. We also discuss how the information about the in-state can be tracked throughout during evolution, and not only at late times, for the CGHS case.

There are many interesting future implications of the ideas we have presented in this paper. Study of such non-vacuum  distortions could be undertaken for more general set ups, which will involve  computation of Bogoliubov coefficients for a more realistic scenario involving angular momentum, charge etc. Even at the vacuum level, there are different sources of non-thermal vacuum response as suggested in \cite{Visser:2014ypa}. A realistic situation encompassing all such distortions will involve Bogoliubov coefficients modified in a precise manner. Therefore, even at the classical level, there should be a more realistic assessment of the allowed symmetry of the initial data viz-\'{a}-viz the information encryption in the distorted spectrum. It is also worthwhile  to study the field configurations corresponding to  states which encode maximal information  and their field theoretic interpretations. 

Apart from dealing with the more generalized situations and the Bogoliubov coefficients therein, a rigorous analysis has to be done for different kinds of initial data. For instance, in this paper our discussion was limited to pure states which are eigenstates of the number operator. Analysis regarding the most general set of initial data --- such as one which is not an eigenstate of the number operator --- will be required to exhaust the full Hilbert space. A classical initial data, after all, might correspond to a coherent state-like  description. Similarly, the analysis could be translated to the language of  wave packets for a realistic physical response in the out-configuration. Further, analogous field theoretic analysis has to be done for a mixed state description, e.g. a thermal initial data.  In these cases, the analysis of higher order correlation will certainly become more important. In any case, we need to study the resulting non thermal spectra from the point of view of various aspects in 
unitary evolution of the black hole such as strength of the correction, first 
bit release time etc.  and its potential to make the evolution unitary in the spirit of \cite{Dvali:2015aja}. We will pursue these aspects elsewhere.

Lastly, a truthful implementation of back-reaction in realistic collapse scenarios in higher dimensions remains due. Recently, 't Hooft has  proposed a model \cite{Hooft:2015jea}, which potentially captures the information about back-reaction through a shift in a null geodesic due to discontinuity across a matter field geodesic. Supplementing the effect of non-vacuum distortion with the shift arising from the back-reaction is expected to reveal more aspects of issues related to information accessible by future asymptotic observers. 
\section*{Acknowledgement}

Research of S.C. is funded by a SPM fellowship from CSIR, Government of India. The research of TP is partially supported by the J.C. Bose research grant of the Department of Science and Technology, Government of India. Authors wish to thank  Sukanta Bose for helpful discussions and useful references.
\appendix 
\labelformat{section}{Appendix #1} 
\labelformat{subsection}{Appendix #1}
\section{Spectrum operator} \label{App01}

Using the expression for the correction term over the vacuum thermal spectrum, we can obtain the distortion from thermal Hawking radiation for one particle initial state of the field which is undergoing the collapse as,
\begin{widetext}
\bea
N_{\Omega}
= \left[
\left|\int_0^{\infty} \frac{d \tilde{\omega}'}{\sqrt{4\pi\tilde{\omega}'}}\alpha^*_{\Omega \tilde{\omega}'}f(\tilde{\omega}')\right|^2+\left|\int_0^{\infty} \frac{d \tilde{\omega}}{\sqrt{4\pi\tilde{\omega}}}\beta_{\Omega \tilde{\omega}}f(\tilde{\omega})\right|^2\right]. \label{NCEx1}
\eea
\end{widetext}
It must be noted that the expression in \ref{NCEx1} is general enough to include cases when the Bogoliubov coefficients as in \ref{BT} are modified by back-reaction, angular momentum, quantum gravity etc. In any case, the non-vacuum part of the radiation spectra provides a constraint for $f(\omega)$ in form of \ref{NCEx1}.
Using \ref{Ftilde}, we can rewrite \ref{NCEx1} as
\bea
N_{\Omega} = \frac{1}{4 \pi}\frac{1}{4 \pi \kappa} \frac{1}{\sinh{\frac{\pi\Omega}{\kappa}}}\left[ \left|\tilde{F}\left(\frac{\Omega}{\kappa}\right)\right|^2+\left|\tilde{F}\left(-\frac{\Omega}{\kappa}\right)\right|^2\right]. \label{NCEx2}
\eea
We can decompose $|\tilde{F}\left(\Omega/\kappa\right)|^2 $ into  symmetric $\tilde{S}\left(\Omega/\kappa\right)$ and an anti-symmetric part $\tilde{A}\left(\Omega/\kappa\right)$
\bea
\left|\tilde{F}\left(\frac{\Omega}{\kappa}\right)\right|^2 = \tilde{S}\left(\frac{\Omega}{\kappa}\right) + \tilde{A}\left(\frac{\Omega}{\kappa}\right).
\eea
With this decomposition, we realize from \ref{NCEx2} that the symmetric part of $|\tilde{F}\left(\Omega/\kappa\right)|^2 $ is entirely characterized by the distribution function $N_{\Omega}$ of the radiation,
\bea
\tilde{S}\left(\frac{\Omega}{\kappa}\right)= 8 \pi^2 \kappa N_{\Omega} \sinh{\frac{\pi\Omega}{\kappa}}.\label{symmetricpart}
\eea
Further, if the in-state is normalized to unity, we have
\bea
\int_{-\infty}^{\infty}d \left(\frac{\Omega}{\kappa} \right) e^{-\frac{\pi\Omega}{\kappa}} \left[\tilde{S}\left(\frac{\Omega}{\kappa}\right) + \tilde{A}\left(\frac{\Omega}{\kappa}\right)\right] = 8 \pi^2, \label{NormConstraint1}
\eea
which together with \ref{symmetricpart} regulates the integral (and hence the asymptotic behavior) of $\tilde{A}\left(\kappa\right)$. Apart from this constraint, $\tilde{A}\left(\kappa\right)$ is a completely arbitrary anti-symmetric function. Therefore, the radiation spectra fixes the symmetric part of the probability density in the Fourier space corresponding to $z$. However, the anti-symmetric part of this probability density remains largely unspecified.

In terms of the function $g(z)$ defined in \ref{Def_gz}, the symmetric part $\tilde{S}\left(\frac{\Omega}{\kappa}\right) $ can be written as
\begin{widetext}
\bea
e^{\pi\frac{\Omega}{\kappa}}\int_{-\infty}^{\infty} d y \int_{-\infty}^{\infty} d z g(z-y/2)g^*(z+y/2) e^{-i\frac{\Omega}{\kappa} y} +
e^{-\pi\frac{\Omega}{\kappa}}\int_{-\infty}^{\infty} d y \int_{-\infty}^{\infty} d z g(z-y/2)g^*(z+y/2) e^{i\frac{\Omega}{\kappa} y } = 2 \tilde{S}\left(\frac{\Omega}{\kappa}\right). 
\eea. 
\end{widetext}
As, we see that $F(\kappa)$ is ``momentum space representation" conjugate to $g(z)$, the above expression can be written in terms of the Wigner function corresponding to the phase space of $(z, \Omega/\kappa)$,
\begin{widetext}
\bea
e^{\pi\frac{\Omega}{\kappa}}\int_{-\infty}^{\infty} d z {\cal W}_{g}\left(z,\frac{\Omega}{\kappa}\right) +
e^{-\pi\frac{\Omega}{\kappa}}\int_{-\infty}^{\infty} d z {\cal W}_{g}\left(z,-\frac{\Omega}{\kappa}\right) = 2 \tilde{S}\left(\frac{\Omega}{\kappa}\right),
\eea
\end{widetext}
where the Wigner function is defined as, 
\bea
{\cal W}_{g}\left(z,\frac{\Omega}{\kappa}\right) =\int_{-\infty}^{\infty} d y g(z-y/2)g^*(z+y/2) e^{-i\frac{\Omega}{\kappa} y}.
\eea
Also, with the relation
\bea
\left|F\left(\frac{\Omega}{\kappa}\right)\right|^2 = \int_{-\infty}^{\infty} d z {\cal W}_{g}\left(z,\frac{\Omega}{\kappa}\right),
\eea
we obtain,
\bea
e^{\pi\frac{\Omega}{\kappa}}\left|F\left(\frac{\Omega}{\kappa}\right)\right|^2 + e^{-\pi\frac{\Omega}{\kappa}}\left|F\left(-\frac{\Omega}{\kappa}\right)\right|^2 = 2 \tilde{S}\left(\frac{\Omega}{\kappa}\right), \label{OddEven}
\eea
which is an obvious illustration of \ref{symmetricpart}.
Therefore, integrating the relation \ref{OddEven} over the frequency range at ${\cal J}^{+}$ we obtain the relation
\begin{widetext}
\bea
2 \int_0^{\infty} d\left(\frac{\Omega}{\kappa}\right) \tilde{S}\left(\frac{\Omega}{\kappa}\right) &=& \int_{0}^{\infty}d \left(\frac{\Omega}{\kappa}\right)e^{\frac{\pi\Omega}{\kappa}} \int_{-\infty}^{\infty} d z {\cal W}_{g}\left(z,\frac{\Omega}{\kappa}\right)+\int_{0}^{\infty}d \left(\frac{\Omega}{\kappa}\right)e^{-\frac{\pi\Omega}{\kappa}} \int_{-\infty}^{\infty} d z {\cal W}_{g}\left(z,-\frac{\Omega}{\kappa}\right) \nonumber\\
&=& \int_{-\infty}^{\infty}d y e^{\pi y}\left|F\left(y\right)\right|^2.
\eea
\end{widetext}
Although, the state which would be completely specified, if we know $F(y)$, remains arbitrary apart form this constraint, the symmetric part $\tilde{S}\left(\Omega/\kappa\right)$ which is completely specified through the non-vacuum distortion, fixes the expectation of the exponentiated momenta conjugate to $z (=\log{\omega/C})$.  In fact one can show (see \ref{MultiParticle}) that this constraint is present not only for a single excitation, but a general $n$-th excited state as well. 

For a n-particle state
\bea \label{nPS}
|\Psi\rangle= \int_0^{\infty}\prod_{i=1}^n \frac{d \omega_i}{\sqrt{2\pi\omega_i}} f(\omega_1,...\omega_n) \hat{a}^{\dagger}(\omega_i)|0 \rangle_M, 
\eea
the radiation profile over the thermal component fixes the expectation of a single particle exponentiated momentum, i.e.
\bea
16  \int_0^{\infty} d\left(\frac{\Omega}{\kappa}\right) \pi^2 \kappa \left(\sinh{\frac{\pi\Omega}{\kappa}}\right)N_{\Omega} =  \frac{1}{n}\langle \Psi | e^{\frac{\pi \hat{\Omega}}{\kappa}}|\Psi\rangle. \label{MultiExpectation}
\eea

Additional information about the initial state can only be obtained from the spectrum if the initial state has some symmetries. We will discuss few interesting cases below.
\begin{itemize}

\item If $F(y)$ is a real and symmetric function, then we see from \ref{OddEven} that it gets completely specified in terms of $\tilde{S}(\Omega/\kappa)$. As, a result the initial state also gets completely specified as $g(z)$ can be obtained by the inverse Fourier transform. However, by virtue of the properties of Fourier transform, $g(z)$ also happens to be real and symmetric. This symmetry corresponds to a duality in the frequency space distribution about the surface gravity parameter $\kappa$. These states are very special class of initial states whose information get coded entirely in the radiation from the black hole within the framework of standard unitary quantum mechanics.

\item For a slightly more general case, the reality condition on $g(z)$ can be traded for by imposing relation between $F(\Omega/\kappa)$ and 
$F(-\Omega/\kappa)$, which is to specify the symmetry of $F(y)$ in the positive and negative half planes. Such a specification of symmetry constrains the distribution $F(y)$ to remain arbitrary in one of the half planes and amounts to reducing the degrees of freedom by half. Let us assume $F(\Omega/\kappa)$ is real, that means
\bea
g(z)=g^*(-z). \label{RealityCondition}
\eea
Now additionally if we impose, 
\bea
F\left(-y\right) = K\left(y\right) F\left(y\right),
\eea
for a specified function $K(y)$, then
\bea
\int_{-\infty}^{\infty}d z g(z) e^{-i y z } = \int_{-\infty}^{\infty}d z g(z) e^{i y z} K\left(y\right). \label{SymmCond}
\eea
Therefore, using the condition \ref{RealityCondition}, we can obtain from \ref{SymmCond}
\bea
g^*(z) &=& \int_{-\infty}^{\infty}d z' g(z') \int_{-\infty}^{\infty}d y K\left(y\right)e^{iy (z'-z)} \nonumber\\
&=& \int_{-\infty}^{\infty}d z' g(z') \tilde{K} (z'-z),
\eea
where $\tilde{K}(q)$ is the inverse Fourier transform of $K(y).$ Therefore, for such a symmetry in the the probability amplitude, the state can be recovered from 
\bea 
F^2\left(\frac{\Omega}{\kappa} \right) = 16 \pi^2 \kappa \frac{\left(\sinh{\frac{\pi\Omega}{\kappa}}\right)}{e^{\frac{\pi\Omega}{\kappa}}+\left(K\left(\frac{\Omega}{\kappa}\right)\right)^2e^{-\frac{\pi\Omega}{\kappa}} } N_{\Omega}.  \label{F4GenrlSym}
\eea

\end{itemize}
Therefore, we see that the symmetry of the prescribed class for one particle sate encodes the entire information of the in-state in the resulting radiation from the black hole. If the initial condition of the collapse demands symmetry of such kinds, the resulting mixed state has enough information in the spectra to completely specify the state. We will further consider some other classes of symmetries in the initial data for spherically symmetric collapse models and their imprints in the non-vacuum distortions. 
\section{Real initial distribution}\label{App_02}

For real distributions, the Fourier transform will satisfy
\bea
|F(y)|^2 = |F(-y)|^2
\eea
Therefore, $|K(y)|=1$ and we have the relation
\bea 
\left|F\left(\frac{\Omega}{\kappa} \right)\right|^2+\left|F\left(-\frac{\Omega}{\kappa} \right)\right|^2 = 8 \pi^2 \kappa \tanh{\frac{\pi\Omega}{\kappa}}N_{\Omega}. 
\eea
For a symmetric algebraic operator of $y$
\bea
\hat{\cal O}_{\text{even}}(y)=\hat{\cal O}_{\text{even}}(-y),
\eea
the expression
\bea
8 \pi^2 \kappa \tanh{\frac{\pi\Omega}{\kappa}}{\cal O}_{\text{even}}\left(\frac{\Omega}{\kappa} \right)N_{\Omega} = \nonumber\\
{\cal O}_{\text{even}}\left(\frac{\Omega}{\kappa} \right)\left|F\left(\frac{\Omega}{\kappa} \right)\right|^2 + {\cal O}_{\text{even}}\left(-\frac{\Omega}{\kappa} \right)\left|F\left(-\frac{\Omega}{\kappa} \right)\right|^2
\eea
which on integration over the whole frequency range gives the expectation value of the operator
\begin{widetext}
\bea
\int_{0}^{\infty} d\left(\frac{\Omega}{\kappa} \right) 8 \pi^2 \kappa \tanh{\frac{\pi\Omega}{\kappa}}{\cal O}_{\text{even}}\left(\frac{\Omega}{\kappa} \right)N_{\Omega} 
 &=& \int_{0}^{\infty} d\left(\frac{\Omega}{\kappa} \right)\left[{\cal O}_{\text{even}}\left(\frac{\Omega}{\kappa} \right)\left|F\left(\frac{\Omega}{\kappa} \right)\right|^2 + {\cal O}_{\text{even}}\left(-\frac{\Omega}{\kappa} \right)\left|F\left(-\frac{\Omega}{\kappa} \right)\right|^2\right] \nonumber\\
&=&\int_{-\infty}^{\infty}{\cal O}_{\text{even}}(y)|F(y)|^2. \label{SymOpEx}
\eea
\end{widetext}
The analysis can also be directly done as derivative operators on frequency representation (see \ref{AppC}).
By similar logic, one can argue that expectation of all odd algebraic operators vanish in this case, i.e. with a symmetric $|F(y)|^2,$
the expectation value for an odd observable
\bea
\langle \hat{\cal O}_{\text{odd}}(y) \rangle = \int _{-\infty}^{\infty} dy {\cal O}_{\text{odd}}(y)|F(y)|^2 =0.
\eea
Thus in this scenario, expectation of all algebraic operators in $y$ will be given in terms of spectral distortion. Any general operator
$\hat{\cal O}(y)$ can be decomposed in terms of its even and odd parts
\bea 
\hat{\cal O}(y) =\hat{\cal O}_{\text{even}}(y) + \hat{\cal O}_{\text{odd}}(y).
\eea
Therefore, to obtain $\langle \hat{\cal O}(y) \rangle$ one only requires $\langle \hat{\cal O}_{\text{even}}(y) \rangle$, which can be easily obtained from \ref{SymOpEx}.
Similarly for the generalized symmetry class
\bea
\langle \hat{\cal O}(y) \rangle  = \int_{0}^{\infty} dy\left[{\cal O}\left(y \right)\left|F\left(y \right)\right|^2 + {\cal O}\left(-y \right)\left|F\left(-y \right)\right|^2\right] \nonumber\\
 = \int_{0}^{\infty} dy({\cal O}\left(y \right)+|K(y)|^2{\cal O}\left(-y \right))\left|F\left(y \right)\right|^2\nonumber\\
 =16 \pi^2 \kappa \int_{0}^{\infty} d\bar{\Omega} \frac{[{\cal O}\left(\bar{\Omega} \right)+|K(\bar{\Omega})|^2{\cal O}\left(-\bar{\Omega} \right)]\sinh{\bar{\Omega}}}{e^{\frac{\pi\Omega}{\kappa}}+\left|K\left(\bar{\Omega}\right)\right|^2e^{-\frac{\pi\Omega}{\kappa}} } N_{\Omega},\nonumber\\
\eea
where we have used the expression of $\left|F\left(y \right)\right|^2$ in the range $y \in(0,\infty)$ from \ref{F4GenrlSym}, in the third equality. Thus even with the specified symmetry class $K(y)$, all the algebraic operators on the momentum space become fixed.

For a general multi-particle state, the above identities provide information regarding the single-particle sector of the state (refer to \ref{MultiParticle}). Since spectral distortion is just one function, it presumably provides information about one sector of the field. Higher correlations of the out-state are expected to contain more such information about the in-state. However, we defer the analysis of such higher correlations to a future work.
\section{Partial information for multiply excited state}\label{App_03}

As argued in the previous section, for many particle state \ref{nPS}, the expression \ref{MultiExpectation} provides the expectation of exponentiated momenta per particle. In addition, as before, specification of additional symmetries provide additional operators' expectations.

For a multi-particle state with above-mentioned symmetries in \ref{App_02}, the spectral distortion completely characterizes the one particle sector of the field. A special multi-particle state of type of this class
\bea
|\Psi\rangle = \int... \int\left(\prod_i d \omega_i \psi(\omega_i)\hat{a}^{\dagger}(\omega_i) \right)|0\rangle_{in},
\eea
with the symmetries discussed above, can be entirely retrieved from the resulting radiation spectra. This state is {\it analogous to a Bose-Einstein condensate} of the particles of the scalar field.

Similarly analogous to the single particle setting, real distributions fix operators on single particle sector of the field.  
We first obtain an identity for a single excitation state, but soon the result will be shown to be applicable on higher excited state as well, which yields the particle content of the multi-particle state. For a real $f(\tilde{\omega})$ and hence real $g(z)= f(C e^z)$,
$$\left |F\left(\frac{\Omega}{\kappa}\right)\right|^2=\left |\int_{-\infty}^{\infty} dt g(z) e^{i y z}\big|_{y=\frac{\Omega}{\kappa}}\right|^2 $$
will be symmetric in $y=\Omega/\kappa$. As a result,
\bea
N_{\Omega}
=\frac{1}{4 \pi}\frac{1}{2 \pi \kappa} \coth{\frac{\pi\Omega}{\kappa}}\left|\int_{-\infty}^{\infty} dz g(z) e^{i \frac{\Omega}{\kappa}z} \right|^2.
\eea
Therefore,
\bea
\int_{-\infty}^{\infty} d\left(\frac{\Omega}{\kappa}\right) N_{\Omega} \tanh{\frac{\pi\Omega}{\kappa}} &=& \frac{1}{4 \pi}\int_{-\infty}^{\infty} dz |g(z)|^2  \nonumber\\
&=& {}_{in}\langle \Psi | \Psi \rangle_{in},
\eea
for the initial state $| \Psi \rangle_{in}$ of the field as given in \ref{state2}.

For a general $n-$particle state \ref{nPS},
with $f(\omega_1,...\omega_n) $ being a real distribution, we can obtain the identity
\bea
2\int_{0}^{\infty} d \left(\frac{\Omega}{\kappa}\right)N_{\Omega}\tanh{\frac{\pi\Omega}{\kappa}}=n\times {}_{in}\langle \Psi | \Psi \rangle_{in}. \label{NExpectationGS}
\eea
Therefore, for real distributions the correction over thermality in the out-state carries the information about number of excitations in the in-state. In fact, this identity is true if the function
\bea
|{\cal F}(y)|^2 =\int_{-\infty}^{\infty}\int_{-\infty}^{\infty}dt dt' \rho_R(t,t') e^{iy(t-t')},
\eea
with $\rho_R(t,t')$ being the reduced one particle density matrix constructed from \ref{nPS} (refer to \ref{AppB}), is symmetric in $y.$  

Within this generalized class of symmetric states the excitation number in the in-state is obtained from \ref{NExpectationGS}.
For this identity to hold we only require
\bea
\frac{F\left(\frac{\Omega}{\kappa}\right)}{F\left(-\frac{\Omega}{\kappa}\right)}=e^{i \alpha_{\Omega}} \hspace{0.25 cm} \text{or,} \hspace{0.25 cm}
\frac{F^*\left(\frac{\Omega}{\kappa}\right)}{F\left(-\frac{\Omega}{\kappa}\right)}=e^{-i \beta_{\Omega}},
\eea
for some arbitrary functions $\alpha_{\Omega}$ and $\beta_{\Omega}$. Real amplitudes happens to be only a subset of this class of conditions. The results of stimulated emissions in the excited state $|n_0,n_1,...\rangle$ can be derived from this.
Within this generalized class of symmetric states the excitation number in the in-state is obtained form the cumulative number expectation value given in \ref{NExpectationGS}. Further a similar kind of identity can be obtained from the symmetry class 
of the kind
\bea
\frac{F\left(\frac{\Omega}{\kappa}\right)}{F\left(-\frac{\Omega}{\kappa}\right)} &=& e^{i \alpha_{\Omega}}K\left(\frac{\Omega}{\kappa}\right) \hspace{0.25 cm} \text{or,} \nonumber\\
\frac{F^*\left(\frac{\Omega}{\kappa}\right)}{F\left(-\frac{\Omega}{\kappa}\right)} &=& e^{-i \beta_{\Omega}}K\left(\frac{\Omega}{\kappa}\right),
\eea
for a specified $K(\Omega/\kappa)$. In that case, the number expectation in the in-state is obtained from the expression
\bea
n=2\times 2\int_{0}^{\infty} d \left(\frac{\Omega}{\kappa}\right) \kappa \frac{\left(\sinh{\frac{\pi\Omega}{\kappa}}\right)}{e^{\frac{\pi \Omega}{\kappa}}+\left|K\left(\frac{\Omega}{\kappa}\right)\right|^2e^{-\frac{\pi\Omega}{\kappa}} }N_{\Omega}.
\eea
Therefore, we have shown that the radiation spectra (being a real profile) fixes one of the parameter related to the initial state. This parameter can be written as an exponentiated operator on the Hilbert space of states of the theory at the onset of collapse. However, in a general case, the state will not be entirely classified in terms of just one single parameter. We also show that there exist a large class of initial one particle states with specified symmetries which have their entire information content imprinted in the outgoing radiation. Apart from this,
for higher excited states, the expectation value of the exponentiated one particle momentum (conjugate to logarithmic energy) always gets determined. The full information for such states is not obtained from the radiation profile. However, yet there exists a class of symmetries, which fixes the total  particle content of the initial state.
\section{Multi-particle states}\label{MultiParticle}

We show that one particle sector of an $n-$ particle configuration of field can be fixed by the non-vacuum spectral distortion, similar to the case of a single particle state. We will demonstrate the case for a two particle state explicitly, but a simple generalization to an $n-$ particle state \ref{nPS} will follow identically.
\subsection{Two Particle state}
A two particle state is specified using the distribution $f(\omega_1, \omega_2) $ in the frequency representation as,
\bea
|\Psi\rangle = \int \frac{d \omega_1}{\sqrt{2\pi}}\frac{d \omega_2}{\sqrt{2\pi}}\frac{f(\omega_1, \omega_2)}{\sqrt{2\omega_12\omega_2}} \hat{a}^{\dagger}_{\omega_1}\hat{a}^{\dagger}_{\omega_2}|0\rangle. \label{2PS}
\eea
The norm of the state is given as
\begin{widetext}
\bea
\langle \Psi| \Psi \rangle =  \int \frac{d \omega_1}{\sqrt{2\pi}}\frac{d \omega_2}{\sqrt{2\pi}}\frac{d \omega_1'}{\sqrt{2\pi}}\frac{d \omega_2'}{\sqrt{2\pi}} \frac{f(\omega_1, \omega_2)}{\sqrt{2\omega_12\omega_2}}\frac{f^*(\omega_1', \omega_2')}{\sqrt{2\omega_1'2\omega_2'}}\times 
\langle 0|\hat{a}_{\omega_1'}\hat{a}_{\omega_2'}\hat{a}^{\dagger}_{\omega_1}\hat{a}^{\dagger}_{\omega_2} |0 \rangle,\nonumber\\
=\frac{1}{4\pi^2} \int \frac{d \omega_1}{\omega_1}\frac{d \omega_2}{\omega_2}[f(\omega_1, \omega_2)f^*(\omega_1, \omega_2)+f(\omega_1, \omega_2)f^*(\omega_2, \omega_1)].\nonumber\\
\eea
For immediate usage, we also evaluate the expression for 
\bea 
\langle \Psi|\hat{a}^{\dagger}_{\omega} \hat{a}_{\omega'}|\Psi \rangle =\frac{1}{\sqrt{4 \pi \omega}}\frac{1}{\sqrt{4 \pi \omega'}}
\int\frac{d \tilde{\omega}}{4\pi\tilde{\omega}}[f(\omega', \tilde{\omega})f^*(\omega, \tilde{\omega})+
f(\tilde{\omega}, \omega')f^*(\omega, \tilde{\omega})+f(\omega',\tilde{\omega})f^*( \tilde{\omega},\omega)+f(\tilde{\omega}, \omega')f^*(\omega, \tilde{\omega})]. \nonumber\\ \label{NEx1term}
\eea 
\end{widetext}
The expression for the non-vacuum distortion to the Hawking radiation will be obtained from 
\begin{widetext}
\bea
N_{\Omega} =
\int_0^{\infty} d \tilde{\omega}\int_0^{\infty} d \tilde{\omega}'
[
\alpha_{\Omega \tilde{\omega}}\alpha^*_{\Omega \tilde{\omega}'}\langle\Psi|\hat{a}^{\dagger}(\tilde{\omega})\hat{a}(\tilde{\omega}')|\Psi\rangle+
\beta_{\Omega \tilde{\omega}}\beta^*_{\Omega \tilde{\omega}'}\langle\Psi|\hat{a}^{\dagger}(\tilde{\omega}')\hat{a}(\tilde{\omega})|\Psi\rangle \nonumber\\
-
\alpha_{\Omega \tilde{\omega}}\beta^*_{\Omega \tilde{\omega}'}\langle\Psi|\hat{a}^{\dagger}(\tilde{\omega})\hat{a}^{\dagger}(\tilde{\omega}')|\Psi\rangle-
\beta_{\Omega \tilde{\omega}}\alpha^*_{\Omega \tilde{\omega}'}\langle\Psi|\hat{a}(\tilde{\omega})\hat{a}(\tilde{\omega}')|\Psi\rangle
].\label{NEx3}
\eea
\end{widetext}
For pure number operator eigenstates expressions $\langle\Psi|\hat{a}^{\dagger}(\tilde{\omega})\hat{a}^{\dagger}(\tilde{\omega}')|\Psi\rangle$ and $\langle\Psi|\hat{a}(\tilde{\omega})\hat{a}(\tilde{\omega}')|\Psi\rangle$ will vanish.
We evaluate the first term in the expression \ref{NEx3} using \ref{NEx1term}.  The second term will be obtained from the exchange $\tilde{\omega}\longleftrightarrow \tilde{\omega}'.$ Thus, for the first term in \ref{NEx1term} first two terms of \ref{NEx3} can be written as

\begin{widetext}
\begin{align}
\int_0^{\infty} d \omega\int_0^{\infty} d \omega' \alpha_{\Omega \omega}\alpha^*_{\Omega \omega'}\langle\Psi|\hat{a}^{\dagger}(\omega)\hat{a}(\omega')|\Psi\rangle = 
\int_0^{\infty} \int_0^{\infty}   \frac{d \omega}{\sqrt{4 \pi \omega}}\frac{d \omega'}{\sqrt{4 \pi \omega'}}
\int_0^{\infty}\frac{d \tilde{\omega}}{4\pi\tilde{\omega}}\alpha_{\Omega \omega}\alpha^*_{\Omega \omega'}f(\omega', \tilde{\omega})f^*(\omega, \tilde{\omega}).
\end{align}

\bea
\int_0^{\infty} d \omega\int_0^{\infty} d \omega' \beta_{\Omega \omega}\beta^*_{\Omega \omega'}\langle\Psi|\hat{a}^{\dagger}(\omega')\hat{a}(\omega)|\Psi\rangle = 
\int_0^{\infty} \int_0^{\infty} \frac{d \omega}{\sqrt{4 \pi \omega}}\frac{d \omega'}{\sqrt{4 \pi \omega'}}
\int_0^{\infty}\frac{d \tilde{\omega}}{4\pi\tilde{\omega}}\beta_{\Omega \omega}\beta^*_{\Omega \omega'}f(\omega, \tilde{\omega})f^*(\omega', \tilde{\omega}).
\eea

Again using \ref{BT} and \ref{vartrans} we can write the above expressions as

\bea
\int_0^{\infty} \int_0^{\infty}   \frac{d \omega}{\sqrt{4 \pi \omega}}\frac{d \omega'}{\sqrt{4 \pi \omega'}}
\int_0^{\infty}\frac{d \tilde{\omega}}{4\pi\tilde{\omega}}\alpha_{\Omega \omega}\alpha^*_{\Omega \omega'}f(\omega', \tilde{\omega})f^*(\omega, \tilde{\omega}) = \nonumber\\
\frac{e^{\pi \bar{\Omega}}}{4\pi \kappa \sinh{\left(\pi \bar{\Omega} \right)}}\int_{-\infty}^{\infty}\frac{d \tilde{t}}{4 \pi}\int_{-\infty}^{\infty}\frac{d t}{\sqrt{4 \pi}}
\int_{-\infty}^{\infty}\frac{d t'}{\sqrt{4 \pi}} e^{-i\bar{\Omega}(t-t')}g^*(t,\tilde{t})g(t',\tilde{t}),
\eea
and,
\bea
\int_0^{\infty} \int_0^{\infty}   \frac{d \omega}{\sqrt{4 \pi \omega}}\frac{d \omega'}{\sqrt{4 \pi \omega'}}
\int_0^{\infty}\frac{d \tilde{\omega}}{4\pi\tilde{\omega}}\beta_{\Omega \omega}\beta^*_{\Omega \omega'}f(\omega', \tilde{\omega})f^*(\omega, \tilde{\omega}) = \nonumber\\
\frac{e^{-\pi \bar{\Omega}}}{4\pi \kappa \sinh{\left(\pi \bar{\Omega} \right)}}\int_{-\infty}^{\infty}\frac{d \tilde{t}}{4 \pi}\int_{-\infty}^{\infty}\frac{d t}{\sqrt{4 \pi}}
\int_{-\infty}^{\infty}\frac{d t'}{\sqrt{4 \pi}} e^{i\bar{\Omega}(t-t')}g^*(t,\tilde{t})g(t',\tilde{t}).
\eea

Adding these two terms we obtain the expression of the first term of  non-vacuum distortion \ref{NEx3}, in the Hawking radiation,

\bea
16 \pi^2 \kappa \sinh{\left(\pi \bar{\Omega} \right)} N_{\bar{\Omega}}\big|_{\text{1st term}}=
e^{\pi \bar{\Omega}}\int_{-\infty}^{\infty}\frac{d \tilde{t}}{4 \pi}\int_{-\infty}^{\infty}d t
\int_{-\infty}^{\infty}d t' e^{-i\bar{\Omega}(t-t')}g^*(t,\tilde{t})g(t',\tilde{t})+\nonumber\\
e^{-\pi \bar{\Omega}}\int_{-\infty}^{\infty}\frac{d \tilde{t}}{4 \pi}\int_{-\infty}^{\infty}d t
\int_{-\infty}^{\infty}d t' e^{i\bar{\Omega}(t-t')}g^*(t,\tilde{t})g(t',\tilde{t}).
\eea
Similarly, remaining terms in \ref{NEx3} can be written as
\bea
16 \pi^2 \kappa \sinh{\left(\pi \bar{\Omega} \right)} N_{\bar{\Omega}}\big|_{\text{2nd term}}=
e^{\pi \bar{\Omega}}\int_{-\infty}^{\infty}\frac{d \tilde{t}}{4 \pi}\int_{-\infty}^{\infty}d t
\int_{-\infty}^{\infty}d t' e^{-i\bar{\Omega}(t-t')}g^*(t,\tilde{t})g(t',\tilde{t})+\nonumber\\
e^{-\pi \bar{\Omega}}\int_{-\infty}^{\infty}\frac{d \tilde{t}}{4 \pi}\int_{-\infty}^{\infty}d t
\int_{-\infty}^{\infty}d t' e^{i\bar{\Omega}(t-t')}g^*(t,\tilde{t})g(\tilde{t},t'),
\eea
\bea
16 \pi^2 \kappa \sinh{\left(\pi \bar{\Omega} \right)} N_{\bar{\Omega}}\big|_{\text{3rd term}}=
e^{\pi \bar{\Omega}}\int_{-\infty}^{\infty}\frac{d \tilde{t}}{4 \pi}\int_{-\infty}^{\infty}d t
\int_{-\infty}^{\infty}d t' e^{-i\bar{\Omega}(t-t')}g^*(\tilde{t},t)g(t',\tilde{t})+\nonumber\\
e^{-\pi \bar{\Omega}}\int_{-\infty}^{\infty}\frac{d \tilde{t}}{4 \pi}\int_{-\infty}^{\infty}d t
\int_{-\infty}^{\infty}d t' e^{i\bar{\Omega}(t-t')}g^*(\tilde{t},t)g(t',\tilde{t}),
\eea
and,
\bea
16 \pi^2 \kappa \sinh{\left(\pi \bar{\Omega} \right)} N_{\bar{\Omega}}\big|_{\text{4th term}}=
e^{\pi \bar{\Omega}}\int_{-\infty}^{\infty}\frac{d \tilde{t}}{4 \pi}\int_{-\infty}^{\infty}d t
\int_{-\infty}^{\infty}d t' e^{-i\bar{\Omega}(t-t')}g^*(\tilde{t},t)g(\tilde{t},t')+\nonumber\\
e^{-\pi \bar{\Omega}}\int_{-\infty}^{\infty}\frac{d \tilde{t}}{4 \pi}\int_{-\infty}^{\infty}d t
\int_{-\infty}^{\infty}d t' e^{i\bar{\Omega}(t-t')}g^*(\tilde{t},t)g(\tilde{t},t').
\eea
Integrating the LHS of each terms above over $\bar{\Omega}$ we can rewrite the terms in a compact form as
\bea
16 \pi^2 \kappa \int_0^{\infty} d \bar{\Omega} \sinh{\left(\pi \bar{\Omega} \right)} N_{\bar{\Omega}}\big|_{\text{1st term}}=
\int_{-\infty}^{\infty} dy e^{\pi y}\int_{-\infty}^{\infty}\frac{d \tilde{t}}{4 \pi}\int_{-\infty}^{\infty}d t
\int_{-\infty}^{\infty}d t' e^{-i y (t-t')}g^*(t,\tilde{t})g(t',\tilde{t}), \label{Spectra4-2PS1}
\eea
\bea
16 \pi^2 \kappa \int_0^{\infty} d \bar{\Omega} \sinh{\left(\pi \bar{\Omega} \right)} N_{\bar{\Omega}}\big|_{\text{2nd term}}=
\int_{-\infty}^{\infty} dy e^{\pi y}\int_{-\infty}^{\infty}\frac{d \tilde{t}}{4 \pi}\int_{-\infty}^{\infty}d t
\int_{-\infty}^{\infty}d t' e^{-i y(t-t')}g^*(t,\tilde{t})g(t',\tilde{t}),
\eea
\bea
16 \pi^2 \kappa \int_0^{\infty} d \bar{\Omega} \sinh{\left(\pi \bar{\Omega} \right)} N_{\bar{\Omega}}\big|_{\text{3rd term}}=
\int_{-\infty}^{\infty} dy e^{\pi y}\int_{-\infty}^{\infty}\frac{d \tilde{t}}{4 \pi}\int_{-\infty}^{\infty}d t
\int_{-\infty}^{\infty}d t' e^{-i y(t-t')}g^*(\tilde{t},t)g(t',\tilde{t}),
\eea
and
\bea
16 \pi^2 \kappa \int_0^{\infty} d \bar{\Omega} \sinh{\left(\pi \bar{\Omega} \right)} N_{\bar{\Omega}}\big|_{\text{4th term}}=
\int_{-\infty}^{\infty} dy e^{\pi y}\int_{-\infty}^{\infty}\frac{d \tilde{t}}{4 \pi}\int_{-\infty}^{\infty}d t
\int_{-\infty}^{\infty}d t' e^{-i y(t-t')}g^*(\tilde{t},t)g(\tilde{t},t').\label{Spectra4-2PS4}
\eea 
\end{widetext}
Adding all these terms we obtain the expression 
\bea
2\int_0^{\infty}d \bar{\Omega} \sinh{\left(\pi \bar{\Omega} \right)} N_{\bar{\Omega}} = \sum_i \frac{1}{2}\langle \Psi|e^{\pi \hat{y}_i}| \Psi \rangle,
\eea
which is the expectation of $\sum_i \langle \Psi|e^{\pi \hat{y}_i}/2 $, giving average expectation per particle. Similarly, the expression can be generalized to the $n-$ particle state \ref{nPS} as
\bea
2\int_0^{\infty}d \bar{\Omega} \sinh{\left(\pi \bar{\Omega} \right)} N_{\bar{\Omega}} = \sum_i \frac{1}{n}\langle \Psi|e^{\pi \hat{y}_i}| \Psi \rangle.
\eea
\section{Reduced Density matrix for multi-particle state}\label{AppB}

As discussed in \ref{App_03}, the symmetries of reduced density matrix decide the amount of information in the distorted spectra. The reduced density matrix is reduced upto  single particle sector, i.e. for $n-$ particle state, we need to trace over $n-1$ particle states to obtain the reduced density matrix. Here, we will show that using a two-particle state yet again. First, will briefly show a reduced density matrix for a two particle state and show how it is related to the spectral distortion. Thus, the reduced density matrix will provide information about the single particle sector of the field. For higher particle states, the procedure can be repeated exactly analogously without ambiguity. Therefore, given a two particle state \ref{2PS}, the density matrix
\begin{widetext}
\bea
\hat{\rho} = |\Psi \rangle \langle \Psi | = \int \frac{d \omega_1}{\sqrt{2\pi}}\frac{d \omega_2}{\sqrt{2\pi}}\int \frac{d \omega_1'}{\sqrt{2\pi}}\frac{d \omega_2'}{\sqrt{2\pi}}\frac{f(\omega_1, \omega_2)}{\sqrt{2\omega_12\omega_2}}\frac{f^*(\omega_1', \omega_2')}{\sqrt{2\omega_1'2\omega_2'}} \hat{a}^{\dagger}_{\omega_1}\hat{a}^{\dagger}_{\omega_2}|0\rangle \langle 0|\hat{a}_{\omega_1'}\hat{a}_{\omega_2'},
\eea
which in the frequency space is given as
\bea
\rho(\omega_1,\omega_2;\omega_1',\omega_2')=\frac{\left[f(\omega_1, \omega_2)f^*(\omega_1', \omega_2') + f(\omega_1, \omega_2)f^*(\omega_2', \omega_1') + f(\omega_2, \omega_1)f^*(\omega_1', \omega_2') + f(\omega_2, \omega_1)f^*(\omega_2', \omega_1')\right]}{16 \pi^2\sqrt{\omega_1\omega_2\omega_1'\omega_2'}}.
\eea
The reduced density matrix is obtained as a traced out version over a frequency
\bea
\rho_R(\omega_1,;\omega_1') &=& \int d \omega_2\rho(\omega_1,\omega_2;\omega_1',\omega_2) \nonumber\\
& =& \int \frac{d \omega_2}{\omega_2}\frac{\left[f(\omega_1, \omega_2)f^*(\omega_1', \omega_2) + f(\omega_1, \omega_2)f^*(\omega_2, \omega_1') + f(\omega_2, \omega_1)f^*(\omega_1', \omega_2) + f(\omega_2, \omega_1)f^*(\omega_2, \omega_1')\right]}{16 \pi^2\sqrt{\omega_1\omega_1'}}.\nonumber\\ \label{ReDensM4_2PS}
\eea
\end{widetext}
As can be seen from \ref{Spectra4-2PS1} to \ref{Spectra4-2PS4} the spectra is obtained through this reduced density matrix \ref{ReDensM4_2PS} integrated with the Bogoliubov coefficients \ref{BT}. A similar result can be obtained for the correlation operator \ref{Cor-Op}.
\section{Derivative Operators in Frequency representation}\label{AppC}

With the specification of initial state, we can also be able to obtain the expectations of some derivative operators, depending upon the symmetry of initial profile. We demonstrate one exercise in this regard, with a real distribution $f(\omega)$.
\bea
32 \pi^2 \kappa N_{\Omega} \tanh{(\pi \bar{\Omega})} =\left|\int_{-\infty}^{\infty} g(t) e^{i \bar{\Omega}t} \right|^2+\left|\int_{-\infty}^{\infty} g(t) e^{-i \bar{\Omega}t} \right|^2.\nonumber\\
\eea 
The cosine transform of the distortion carries information about derivative operators as can be seen as follows
\begin{widetext}
\bea
32 \pi^2 \kappa \int_0^{\infty} d\bar{\Omega}\cos{(\pi \bar{\Omega} \epsilon)} N_{\Omega} \tanh{(\pi \bar{\Omega})} 
&=& \int_{-\infty}^{\infty}d t \int_{-\infty}^{\infty}d t'g(t)g^*(t')\{\delta(t-t'+\epsilon)+ \delta(t-t'-\epsilon) \},\nonumber\\
&=& \int_{-\infty}^{\infty}d t [g(t)g^*(t+\epsilon)+g(t)g^*(t-\epsilon)].
\eea 
Therefore, we obtain 
\bea
\lim_{\epsilon \rightarrow 0}\frac{1}{2\epsilon}\frac{d}{d\epsilon}\left[32 \pi^2 \kappa \int_0^{\infty} d\bar{\Omega}\cos{(\pi \bar{\Omega} \epsilon)} N_{\Omega} \tanh{(\pi \bar{\Omega})}\right] = \int_{-\infty}^{\infty}d t g(t)\frac{d^2}{dt^2}g^*(t)=-\langle \Psi |
\hat{y}^2 | \Psi \rangle.
\eea

Clearly all even ordered derivative operators can also be obtained through this mechanism
\bea
\lim_{\epsilon \rightarrow 0}\frac{1}{2\epsilon}\frac{d^{(2n-1)}}{d\epsilon^{(2n-1)}}\left[32 \pi^2 \kappa \int_0^{\infty} d\bar{\Omega}\cos{(\pi \bar{\Omega} \epsilon)} N_{\Omega} \tanh{(\pi \bar{\Omega})}\right] = \int_{-\infty}^{\infty}d t g(t)\frac{d^{(2n)}}{dt^{(2n)}}g^*(t)=(i)^{2n}\langle \Psi |
\hat{y}^{2n} | \Psi \rangle.
\eea
\end{widetext}
Only odd ordered derivative operators remain to be specified, however, for with such an even symmetry of states expectation of odd operators turn out to be vanishing. Thus, all derivative operators in the frequency representation get completely specified.
\section{ State for step function support}\label{Familyofstates}

Let us excite some right-moving modes beyond $x_i^+$ (for simplicity we work with single particle states), such that the normal ordered operator
$\hat{T}_{++}(x^+)$ has support only in the region inside the horizon, i.e.,
\bea
 \langle \hat{T}_{++}(x^+) \rangle_{\text{Regularized}}= h(x^+)\Theta(x^+ - x_i^+), \label{T-Expextation}
\eea
for some well behaved function $ h(x^+)$ and the step function  $\Theta(x^+)$.

If the single particle state is taken to be in the frame of observers which would have described the linear dilaton vacuum, then
\bea
|\Psi \rangle = \int_\omega f(\omega) \hat{a}_{\omega}^{\dagger}|0\rangle,
\eea
where $\int _{\omega}$ stands for $\int d\omega /\sqrt{4\pi\omega}$
and the right-moving quantum field is given on $\mathcal{J}_L^{+}$ as
\bea
\hat{f}_{+}(y^+) = \int_\omega (\hat{a}_{\omega} u_\omega(y^+) + \hat{a}_{\omega}^{\dagger} u^*_\omega(y^+) ), \label{FieldonPast}
\eea
with mode functions $u_\omega(y^+)$. Then the equation \ref{T-Expextation} can be re-written as
\bea
\left|\int_\omega f(\omega)u'_\omega(y^+) \right|^2 = h_1(y^+)\Theta(y^+ - y_i^+), \label{T-Expextation2}
\eea
where $'$ denotes a derivative with respect to $y^+$  and $y_i^+$ marking the location corresponding to $x_i^+$. The function $h_1(y^+)$ absorbs the 
Jacobian of transformation from $x^{\pm}$ basis to $y^{\pm}$ basis,
\bea
T_{++}(x^+)=\frac{\partial y^{\mu}}{\partial x^+}\frac{\partial y^{\nu}}{\partial x^+}T_{\mu \nu}(y^+)=
\frac{\partial y^{+}}{\partial x^+}\frac{\partial y^{+}}{\partial x^+}T_{++}(y^+)
\eea
The condition \ref{T-Expextation2} can be realized by
\bea
\int_\omega f(\omega)u'_\omega(y^+) = \tilde{h}(y^+)\Theta(y^+ - y_i^+),
\eea
with some other well behaved function $\tilde{h}(y^+)$. Owing to the conformal flatness of the two dimensional spacetime and the conformal 
nature of minimally coupled massless scalar field, the mode functions can be written as
\bea
u'_\omega(y^+)=-i\omega u_\omega(y^+).
\eea
Therefore, we only require to have 
\bea
\int_\omega f(\omega)\omega u_\omega(x^+) = \tilde{h}(x^+)\Theta(x^+ - x_i^+) =\zeta(x^+), \label{T-Expextation3}
\eea
where we have absorbed the factor $i$ in the redefinition of $\tilde{h}(x^+)$. Taking the inner product of the \ref{T-Expextation3} with 
itself and using the orthonormal properties of the mode functions, we write 
\bea
\int_\omega \omega^2 |f(\omega)|^2 = (\zeta(x^+),\zeta(x^+)). \label{Condition-CGHS}
\eea
For the states satisfying \ref{Condition-CGHS}, the expression \ref{T-Expextation3} can be inverted for given $\zeta(x^+)$, using the completeness of mode functions to obtain a consistent state. Therefore, the one particle states respecting \ref{Condition-CGHS} will have mode excitations beyond $x_i^+$.
\section{Information Retrieval for the CGHS black hole} \label{App_H}

Using a particular representation of the Gamma function
\bea
\Gamma[z] = i^z \int_{-\infty}^{\infty} d q e^{z q} e^{- i e^q},
\eea
we can write down a product formula
\begin{widetext}
\bea
\Gamma[i(\bar{\omega}-\bar{\omega}')]\Gamma[-i(\bar{\omega}-\bar{\omega}'')] = e^{-\pi \bar{\omega}} e^{\frac{\pi}{2}(\bar{\omega}'+\bar{\omega}'')}\int_{-\infty}^{\infty}d q_1 d q_2 e^{i\bar{\omega}(q_{1}-q_{2})} e^{i(\bar{\omega}'q_1-\bar{\omega}''q_2)}e^{- i e^{q_1} + i e^{q_2}},
\eea
which appears in the spectrum operator expression.
The correction in the vacuum thermal radiation as received by asymptotic left moving observer can be expressed as
\bea
2\pi \lambda \sinh{\pi \bar{\omega}} N_{\bar{\omega}}=\int_0^{\infty}\int_0^{\infty} \frac{d\bar{\omega}'}{\bar{\omega}'}\frac{d\bar{\omega}''}{\bar{\omega}''}\text{Sym}_{\bar{\omega}}\left[\frac{\Gamma[i(\bar{\omega}-\bar{\omega}')]\Gamma[-i(\bar{\omega}-\bar{\omega}'')]}{\Gamma[-i\bar{\omega}']\Gamma[i\bar{\omega}'']} \right]\left|y_i^{+}\right|^{i(\bar{\omega}'-\bar{\omega}'')}f(\bar{\omega}')f^{*}(\bar{\omega}''), \label{CGHS_Spectral_Distortion01}
\eea
where $\text{Sym}_{x}[f(x)] =(f(x)+f(-x))/2$ with $\bar{\omega'} =\omega'/\lambda$.
Using \ref{Trns1}, the spectral distortion can be re-written as,
\bea
2\pi \lambda \sinh{\pi \bar{\omega}} N_{\bar{\omega}}=\int_0^{\infty} d\bar{\omega}'d\bar{\omega}''\int_{-\infty}^{\infty}d q_1 d q_2\text{Sym}_{\bar{\omega}}[e^{-\pi\bar{\omega}} e^{i\bar{\omega}(q_{1}-q_{2})}] e^{i(\bar{\omega}'q_1-\bar{\omega}''q_2)}e^{- i e^{q_1} + i e^{q_2}}g(\omega')g^{*}(\omega'').
\eea
Using, $g(\bar{\omega})$, introduce yet another function
\bea
\chi(q)=e^{-i e^q}\int_0^{\infty} d\bar{\omega} e^{-i \bar{\omega} q} g(\bar{\omega}),\label{Trns2}
\eea
to express the spectral distortion as
\bea
2\pi \lambda \sinh{\pi \bar{\omega}} N_{\bar{\omega}} &=& \int_{-\infty}^{\infty}d q_1 d q_2\text{Sym}_{\bar{\omega}}[e^{-\pi\bar{\omega}} e^{i\bar{\omega}(q_{1}-q_{2})}] \chi(q_1)\chi^*(q_2). \\
4\pi \lambda \sinh{\pi \bar{\omega}} N_{\bar{\omega}} &=& \int_{-\infty}^{\infty}d q_1 d q_2[e^{-\pi\bar{\omega}} e^{i\bar{\omega}(q_{1}-q_{2})} + e^{\pi\bar{\omega}} e^{-i\bar{\omega}(q_{1}-q_{2})}] \chi(q_1)\chi^*(q_2), 
\eea
\end{widetext}
which simply gives 
\bea
4\pi \lambda \sinh{\pi \bar{\omega}} N_{\bar{\omega}}=e^{\pi\bar{\omega}} |{\cal F}_{\chi}(\bar{\omega})|^2 + e^{-\pi\bar{\omega}} |{\cal F}_{\chi}(-\bar{\omega})|^2, \label{CGHS_Spectral_Distortion02}
\eea
with ${\cal F}_{\chi}(\bar{\omega})$ being the Fourier transform of $\chi(q)$ w.r.t. $\bar{\omega}$
\bea
{\cal F}_{\chi}(\bar{\omega}) = \int_{-\infty}^{\infty} d q e^{-i \bar{\omega} q} \chi(q), \label{Trns3}
\eea
which gives \ref{CGHS_Spectral_Distortion02} as the analogue of the \ref{OddEven} for the spherical symmetric collapse. Therefore, we can follow the same steps as outlined in \ref{Sec_02A} and \ref{Sec_03} to recover informations regarding ${\cal F}_{\chi}(\bar{\omega})$. Using the inverse transformations \ref{Trns3}, \ref{Trns2} and \ref{Trns1} we can recover the information regarding the field state $f(\bar{\omega})$ using the moments of ${\cal F}_{\chi}(\bar{\omega})$.
\bibliography{Gravity_1_full,Gravity_2_partial,Brane.bib}

\bibliographystyle{./utphys1}
\end{document}